\documentclass[twocolumn]{aastex63}

\shorttitle{{\it AKARI} sources without HSC counterparts}
\shortauthors{Toba et al.}
\begin{document}

%\title{Infrared Galaxies without Optical Counterparts of Subaru Hyper Suprime-Cam in the AKARI North Ecliptic Pole Wide Survey Field}
\title{Search for Optically Dark Infrared Galaxies without Counterparts of Subaru Hyper Suprime-Cam in the AKARI North Ecliptic Pole Wide Survey Field}

\correspondingauthor{Yoshiki Toba}
\email{toba@kusastro.kyoto-u.ac.jp}

\author[0000-0002-3531-7863]{Yoshiki Toba}
\affiliation{Department of Astronomy, Kyoto University, Kitashirakawa-Oiwake-cho, Sakyo-ku, Kyoto 606-8502, Japan}
\affiliation{Academia Sinica Institute of Astronomy and Astrophysics, 11F of Astronomy-Mathematics Building, AS/NTU, No.1, Section 4, Roosevelt Road, Taipei 10617, Taiwan}
\affiliation{Research Center for Space and Cosmic Evolution, Ehime University, 2-5 Bunkyo-cho, Matsuyama, Ehime 790-8577, Japan}

\author[0000-0002-6821-8669]{Tomotsugu Goto}
\affiliation{Institute of Astronomy, National Tsing Hua University, No. 101, Section 2, Kuang-Fu Road, Hsinchu City 30013, Taiwan}

\author{Nagisa Oi}
\affiliation{Tokyo University of Science, 1-3, Kagurazaka Shinjuku-ku Tokyo 162-8601 Japan}

\author{Ting-Wen Wang}
\affiliation{Institute of Astronomy, National Tsing Hua University, No. 101, Section 2, Kuang-Fu Road, Hsinchu City 30013, Taiwan}

\author[0000-0001-9970-8145]{Seong Jin Kim}
\affiliation{Institute of Astronomy, National Tsing Hua University, No. 101, Section 2, Kuang-Fu Road, Hsinchu City 30013, Taiwan}

\author[0000-0002-8560-3497]{Simon C.-C. Ho}
\affiliation{Institute of Astronomy, National Tsing Hua University, No. 101, Section 2, Kuang-Fu Road, Hsinchu City 30013, Taiwan}

\author[0000-0002-4193-2539]{Denis Burgarella}
\affiliation{Aix Marseille Univ. CNRS, CNES, LAM Marseille, France}

\author{Tetsuya Hashimoto}
\affiliation{Institute of Astronomy, National Tsing Hua University, No. 101, Section 2, Kuang-Fu Road, Hsinchu City 30013, Taiwan}
\affiliation{Centre for Informatics and Computation in Astronomy (CICA), National Tsing Hua University, 101, Section 2, Kuang-Fu Road, Hsinchu, 30013, Taiwan}

\author[0000-0001-5615-4904]{Bau-Ching Hsieh}
\affiliation{Academia Sinica Institute of Astronomy and Astrophysics, 11F of Astronomy-Mathematics Building, AS/NTU, No.1, Section 4, Roosevelt Road, Taipei 10617, Taiwan}

\author[0000-0001-7200-8157]{Ting-Chi Huang}
\affiliation{Department of Space and Astronautical Science,The Graduate University for Advanced Studies, SOKENDAI, 3-1-1 Yoshinodai,Chuo-ku, Sagamihara, Kanagawa 252-5210, Japan}
\affiliation{Institute of Space and Astronautical Science, Japan Aerospace Exploration Agency, 3-1-1 Yoshinodai, Chuo-ku, Sagamihara, Kanagawa 252-5210, Japan}

\author{Ho Seong Hwang}
\affiliation{Korea Astronomy and Space Science Institute, 776 Daedeokdae-ro, Yuseong-gu, Daejeon 34055, Republic of Korea}

\author[/0000-0002-1207-1979]{Hiroyuki Ikeda}
\affiliation{National Astronomical Observatory of Japan, 2-21-1 Osawa, Mitaka, Tokyo 181-8588, Japan}
\affiliation{National Institute of Technology, Wakayama College, Gobo, Wakayama 644-0023, Japan}

\author{Helen K. Kim}
\affiliation{Department of Physics and Astronomy, UCLA, 475 Portola Plaza, Los Angeles, CA 90095-1547, USA}

\author{Seongjae Kim}
\affiliation{University of Science and Technology, Daejeon 34113, Korea}
\affiliation{Korea Astronomy and Space Science Institute, Daejeon 34055, Korea}

\author{Dongseob Lee}
\affiliation{Department of Earth Science Education, Kyungpook National University, Daegu 41566, Korea}

\author[0000-0001-6919-1237]{Matthew A. Malkan}
\affiliation{Department of Physics and Astronomy, UCLA, 475 Portola Plaza, Los Angeles, CA 90095-1547, USA}

\author{Hideo Matsuhara}
\affiliation{Institute of Space and Astronautical Science, Japan Aerospace Exploration Agency, 3-1-1 Yoshinodai, Chuo-ku, Sagamihara, Kanagawa 252-5210, Japan}
\affiliation{Department of Space and Astronautical Science,The Graduate University for Advanced Studies, SOKENDAI, 3-1-1 Yoshinodai,Chuo-ku, Sagamihara, Kanagawa 252-5210, Japan}

\author[0000-0002-7562-485X]{Takamitsu Miyaji}
\affiliation{Instituto de Astronom\'a sede. Ensenada, Universidad Nacional Aut\'onoma de M\'exico (UNAM), Km 107, Carret. Tij.-Ens., Ensenada, 22060, BC, M\'exico}
\affiliation{Leibniz Institut f\"ur Astrophysik Potsdam, An der Sternwarte  16, 14482 Potsdam, Germany\footnote{On sabbatical leave from UNAM.}}

\author[0000-0002-8857-2905]{Rieko Momose}
\affiliation{Department of Astronomy, School of Science, The University of Tokyo, 7-3-1 Hongo, Bunkyo-ku, Tokyo 113-0033, Japan}

\author{Youichi Ohyama}
\affiliation{Academia Sinica Institute of Astronomy and Astrophysics, 11F of Astronomy-Mathematics Building, AS/NTU, No.1, Section 4, Roosevelt Road, Taipei 10617, Taiwan}

\author{Shinki Oyabu}
\affiliation{Institute of Liberal Arts and Sciences, Tokushima University, Minami Jousanjima-Machi 1-1, Tokushima, Tokushima 770-8502, Japan}

\author{Chris Pearson}
\affiliation{RAL Space, STFC Rutherford Appleton Laboratory, Didcot, Oxfordshire, OX11 0QX, UK}
\affiliation{Oxford Astrophysics, University of Oxford, Keble Rd, Oxford OX1 3RH, UK}

\author{Daryl Joe D. Santos}
\affiliation{Institute of Astronomy, National Tsing Hua University, No. 101, Section 2, Kuang-Fu Road, Hsinchu City 30013, Taiwan}

\author{Hyunjin Shim}
\affiliation{Department of Earth Science Education, Kyungpook National University, Daegu 41566, Korea}

\author{Toshinobu Takagi}
\affiliation{Japan Space Forum, 3-2-1, Kandasurugadai, Chiyoda-ku, Tokyo 101-0062, Japan}

\author[0000-0001-7821-6715]{Yoshihiro Ueda}
\affiliation{Department of Astronomy, Kyoto University, Kitashirakawa-Oiwake-cho, Sakyo-ku, Kyoto 606-8502, Japan}

\author[0000-0001-6161-8988]{Yousuke Utsumi}
\affiliation{SLAC National Accelerator Laboratory, 2575 Sand Hill Road, Menlo Park, CA 94025, USA}
\affiliation{Kavli Institute for Particle Astrophysics and Cosmology, Stanford University, 452 Lomita Mall, Stanford, CA 94035, USA}

\author{Takehiko Wada}
\affiliation{Institute of Space and Astronautical Science, Japan Aerospace Exploration Agency, 3-1-1 Yoshinodai, Chuo-ku, Sagamihara, Kanagawa 252-5210, Japan}



\begin{abstract}
We present the physical properties of {\it AKARI} sources without optical counterparts in optical images from the Hyper Suprime-Cam (HSC) on the Subaru telescope.
Using the {\it AKARI} infrared (IR) source catalog and HSC optical catalog, we select 583 objects that do not have HSC counterparts in the {\it AKARI} North Ecliptic Pole (NEP) wide survey field ($\sim 5$ deg$^{2}$).
Because the HSC limiting magnitude is deep ($g_{\rm AB}$ $\sim 28.6$), these are good candidates for extremely red star-forming galaxies (SFGs) and/or active galactic nuclei (AGNs), possibly at high redshifts.
We compile multi-wavelength data out to 500 $\micron$ and use it for Spectral Energy Distribution (SED) fitting with {\tt CIGALE} to investigate the physical properties of {\it AKARI} galaxies without optical counterparts.
We also compare their physical quantities with {\it AKARI} mid-IR selected galaxies with HSC counterparts.
The estimated redshifts of {\it AKARI} objects without HSC counterparts range up to $z\sim 4$,
significantly higher than that of {\it AKARI} objects with HSC counterparts.
We find that: (i) 3.6 $-$ 4.5 $\micron$ color, (ii) AGN luminosity, (iii) stellar mass, (iv) star formation rate, and (v) $V$-band dust attenuation in the interstellar medium of {\it AKARI} objects without HSC counterparts are systematically larger than those of {\it AKARI} objects with counterparts.
These results suggest that our sample includes luminous, heavily dust-obscured SFGs/AGNs at $z\sim 1-4$ that are missed by previous optical surveys, providing very interesting targets for the coming {\it James Webb Space Telescope} era.
\end{abstract}


%%%%%%%%%%%%%%%
% Introduction
%%%%%%%%%%%%%%%
\section{Introduction} 
\label{intro}

In the last two decades, it has become clear that dusty star-forming galaxies (SFGs) and active galactic nuclei (AGNs) play an important role in galaxy formation and evolution, and in co-evolution of galaxies and supermassive black holes (SMBHs) \citep[see e.g.,][and references therein]{Goto11,Casey,Hickox,Chen}.
They are particularly numerous in the high-z universe ($z >2$), the key epoch where the cosmic star formation rate (SFR) density and BH mass accretion rate density reach a maximum \citep[e.g.,][]{Madau,Ueda,Aird}.
\cite{Blecha} conducted a smoothed-particle hydrodynamics and N-body simulation and reported that infrared (IR)  color (3.4 and 4.6 $\micron$ color) correlates with AGN activity and nuclear obscuration in the framework of galaxy merger events--a redder system tends to have large AGN luminosity and hydrogen column density \citep[see also][]{Ricci,Ellison,Gao}.
Since dusty SFGs/AGNs often have redder color in the optical and near-IR (NIR), they could correspond to the maximum phase of SF/AGN activity. They would thus be the crucial population to understanding how the galaxies and their SMBHs co-evolve behind a large amount of dust \citep[see also][]{Hopkins,Narayanan}.

Owing to the heavy extinction by so much dust, some high-z dusty AGNs/SFGs are optically faint or even optically ``dark''.
For example, dust-obscured galaxies \citep[DOGs:][]{Dey,Toba_15,Toba_17,Noboriguchi} are characterized by a large mid-IR (MIR) -- optical color; their optical flux density is about 10$^{3}$ times fainter than that in the MIR band \citep[see also][]{Hwang_13a,Hwang_13b}.
They are considered as dusty SFGs or AGNs at $z \sim$1--2--the relative proportion of AGN/SF could increase with IR luminosity \citep{Melbourne,Toba_17c,Toba_17d,Riguccini}.
A more extreme DOG population, Hot DOGs \citep{Eisenhardt,Wu} known as dusty AGNs at $z > 2$ also tend to be optically-faint.
Recently, some galaxies detected by Atacama Large Millimeter/submillimeter Array (ALMA) are recognized as invisible SFGs without optical/NIR counterparts \citep[e.g.,][]{Franco,Yamaguchi,Williams,Wang}.

{\it AKARI}, the first Japanese space satellite dedicated to IR astronomy \citep{Murakami}, has also a great potential to find such optically dark objects.
In addition to all-sky surveys with MIR and far-IR (FIR) \citep{Ishihara,Yamamura}, {\it AKARI} performed deep and wide observations of the North Ecliptic Pole (NEP) over a total area of 5.4 deg$^{2}$ \citep{Matsuhara}.
The {\it AKARI} NEP survey consists of two layers: NEP Wide \citep[NEP-W;][]{Lee,Kim} and NEP Deep \citep[NEP-D;][]{Wada,Takagi,Murata}\footnote{AKARI NEP-D is a part of NEP-W as shown in Figure \ref{Sky}. But each catalog in NEP-W and NEP-D was independently created. Therefore, we ensure a uniform survey depth even for objects in the NEP-D as long as we use the NEP-W catalog.}.
{\it AKARI} NEP regions were observed with the Infrared Camera \citep[IRC:][]{Onaka}, using its nine filters that continuously cover 2--25 $\micron$.
They are called N2, N3, and N4 for the NIR bands, S7, S9W, and S11 for the shorter part of the MIR band (MIR-S), and L15, L18W, and L24 for the longer part of the MIR bands (MIR-L).
The effective wavelength of the N2, N3, N4, S7, S9W, S11, L15, L18W, and L24 filters is about 2.4, 3.2, 4.1, 7.0, 9.0, 11.0, 15.0, 18.0, and 24.0 $\micron$, respectively. 
These continuous NIR-MIR filters that are critical to trace dust emission heated by AGNs and/or Polycyclic Aromatic Hydrocarbon (PAH) emission associated with SF activity. This makes {\it AKARI}/NEP data quite powerful to select AGNs and measure the SF activity up to $z\sim 2$ \citep[e.g.,][]{Goto11,Murata14,Kevin,Kim19,Poliszczuk}.

Recently, the {\it AKARI} NEP-W has been observed with Hyper Suprime-Cam \citep[HSC; ][]{Miyazaki} \citep
[see also][]{Furusawa,Kawanomoto,Komiyama} on the Subaru telescope (PI: T.Goto).
\cite{Oi} reduced the data and used it to construct a five-band HSC catalog that contains 3,251,792 sources.
The 5$\sigma$ detection limit for $g$-, $r$-, $i$-, $z$-, and $y$-band is about 28.6, 27.3, 26.7, 26.0, and 25.6 mag, respectively.
\cite{Oi} cross-identified the HSC catalog with {\it AKARI} NEP-W catalog and found that $\sim$90,000 {\it AKARI} objects have HSC counterparts (hereafter AKARI--HSC objects) (see also S. J. Kim et al. 2020 in preparation).
Using the AKARI--HSC objects, \cite{Goto} derived their IR luminosity function at $0.35 < z < 2.2$ and measured IR luminosity density as a function of redshift up to $z\sim 2$.
Recently, \cite{Simon} calculated photometric redshifts ($z_{\rm photo}$) of AKARI--HSC objects, and demonstrated how the deep HSC data improve their accuracy.
\cite{Tina} investigated the physical properties of AKARI--HSC objects based on spectral energy distribution (SED) fitting with {\tt CIGALE}\footnote{\url{https://cigale.lam.fr/2018/11/07/version-2018-0/}} \citep[Code Investigating GALaxy Emission:][]{Burgarella,Noll,Boquien} \citep[see also][]{Chiang}.

However, those works focused on {\it AKARI} sources that have optical (i.e., HSC) counterparts.
In this work, we shed light on the remaining objects; {\it AKARI} sources without HSC counterparts in the {\it AKARI} NEP-W.
Since the limiting magnitude of the HSC is deep, these sources are expected to be extremely red SFGs/AGNs at high redshifts. We carefully select the candidates and investigate their physical properties.

The structure of this paper is as follows. Section \ref{DA} describes the data set, sample selection of {\it AKARI} sources without HSC counterparts, and our SED modeling of them.
In Section \ref{R}, we present the results of our SED fitting and the derived physical quantities.
In Section \ref{D}, we compare the resultant physical properties with {\it AKARI} sources that have HSC counterparts \citep{Tina}.
We summarize the results of the work in Section \ref{Sum}.
Throughout this paper, the adopted cosmology is a flat universe with $H_0$ = 70.4 km s$^{-1}$ Mpc$^{-1}$, $\Omega_M$ = 0.272, and $\Omega_{\Lambda}$ = 0.728 \citep[the {\it Wilkinson Microwave
Anisotropy Probe 7} cosmology:][]{Komatsu}, which are the same as those adopted in \cite{Tina}.
Unless otherwise noted, all magnitudes refer to the AB system and a \cite{Salpeter} initial mass function (IMF) is assumed.

%%%%%%%%%%%%%%%%%%%%%%
% DATA and ANALYSIS
%%%%%%%%%%%%%%%%%%%%%%
\section{Data and Analysis}
\label{DA}

%------------------
%    Data set
%------------------
\subsection{Data Set}
\label{DS}

\begin{figure}[h]
    \centering
    \includegraphics[width=0.45\textwidth]{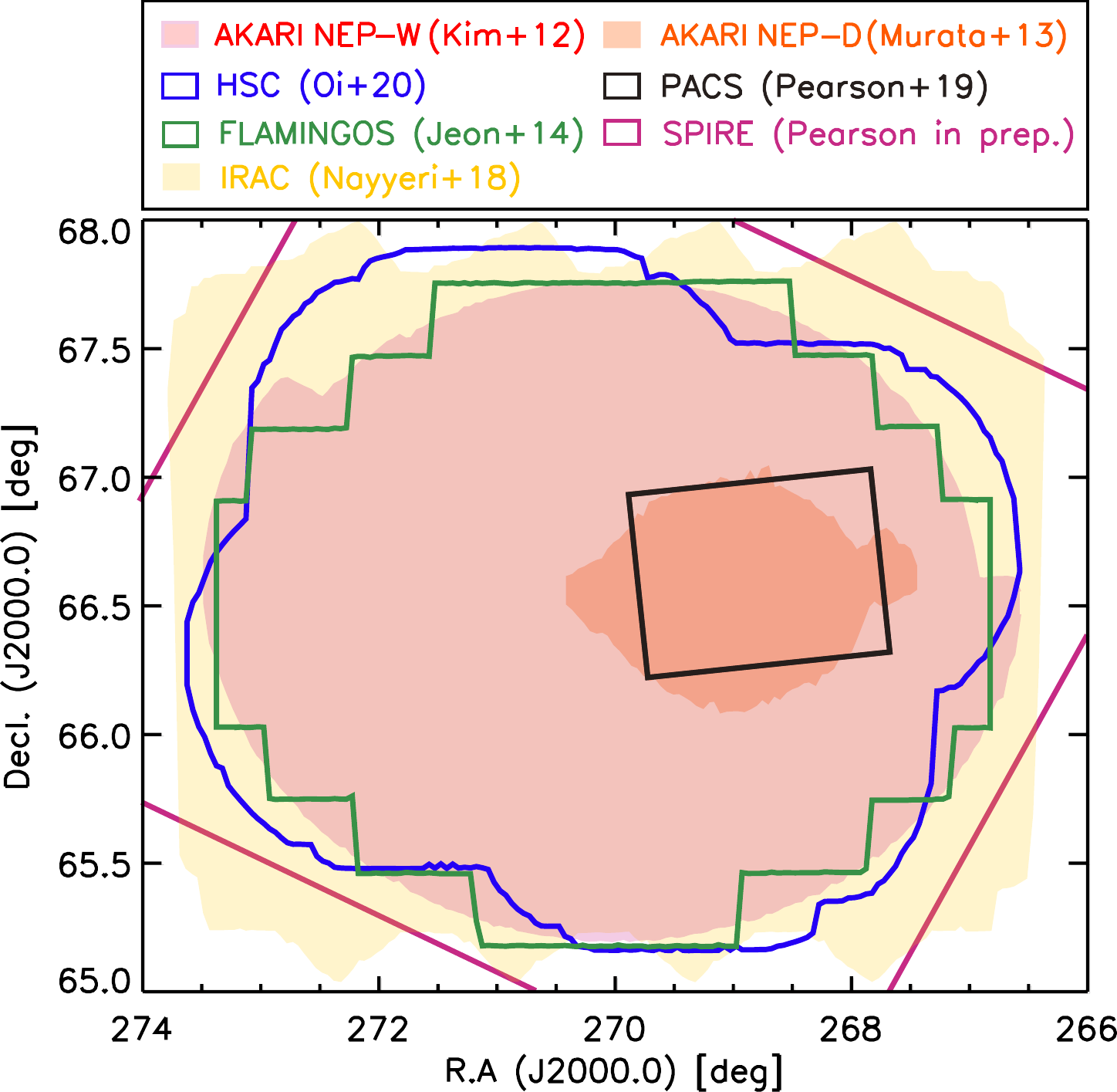}
\caption{The footprint of each observation. Red, orange, and yellow shaded regions represent {\it AKARI} NEP-W \citep{Kim}, NEP-D \citep{Murata}, and IRAC/{\it Spitzer} \citep{Nayyeri}, respectively. Blue, green, black, and magenta line represent HSC/Subaru \citep{Oi}, FLAMINGOS/KPNO \citep{Jeon}, PACS/{\it Herschel} \citep{Pearson}, and SPIRE/{\it Herschel} \citep[][C. Pearson in preparation]{Pearson17}, respectively.}
\label{Sky}
\end{figure}

To select {\it AKARI} sources without HSC counterparts securely and investigate their physical properties, we compile multi-wavelength data. 
{\it AKARI} NEP-W was observed by many facilities, and thus we have abundant dataset from ultraviolet (UV) to radio \citep[see e.g.,][and references therein for a full description]{Kim,Oi}.
We particularly utilized the following data sets in addition to the {\it AKARI} NEP-W catalog \citep{Kim} and HSC catalog \citep{Oi} that are summarized in Table \ref{tbl_DS}.
The area coverage of each dataset is summarized in Figure \ref{Sky}.

\begin{table}
\renewcommand{\thetable}{\arabic{table}}
\centering
\caption{Multi-wavelength Data Set.}
\label{tbl_DS}
\begin{tabular}{ccc}
\tablewidth{0pt}
\hline
\hline
Catalog & Band & 5$\sigma$ detection limit (unit)\\
(Number of sources) &		&		\\
 
\hline
HSC				&	$g$	 	&	28.6 (AB mag)\\
(3,251,792)		&	$r$	 	&	27.3 (AB mag)\\
				&	$i$	 	&	26.7 (AB mag)\\
				&	$z$	 	&	26.0 (AB mag)\\
				&	$y$	 	&	25.6 (AB mag)\\
\hline
FLAMINGOS 		&	$J$		&	21.6 (AB mag)	\\
(295,383)		&	$H$		&	21.3 (AB mag)	\\
\hline
$Spitzer$/IRAC	&	Ch1		&	6.45 ($\mu$Jy)	\\
(380,858)		&	Ch2		&	3.95 ($\mu$Jy)	\\
\hline
$AKARI$ NEP-W 	&	N2		&	15.42 ($\mu$Jy)	\\
(114,794)		&	N3		&	13.30 ($\mu$Jy)\\
				&	N4		&	13.55 ($\mu$Jy)	\\
				&	S7		&	58.61 ($\mu$Jy)	\\
				&	S9W		&	67.30 ($\mu$Jy)	\\
				&	S11		&	93.76 ($\mu$Jy)	\\
				&	L15		&	133.1 ($\mu$Jy)	\\
				&	L18W	&	120.2 ($\mu$Jy)	\\
				&	L24		&	274.4 ($\mu$Jy)	\\
\hline
$Herschel$/SPIRE&	250	 $\micron$	&	9.0 (mJy)	\\
(4,820)			&	350	 $\micron$	&	7.5 (mJy)	\\
				&	500	 $\micron$	&	10.8 (mJy)	\\
\hline
\end{tabular}
\end{table}

We used NIR data provided by \cite{Jeon} who conducted a deep imaging with $J$- and $H$- bands taken by FLoridA Multi-object ImagingNear-ir Grism Observational Spectrometer \citep[FLAMINGOS:][]{Elston} on the Kitt Peak National Observatory (KPNO) 2.1 m telescope.
This photometric catalog contains 295,383 sources with a 5$\sigma$ detection limit of 21.6 and 21.3 mag in the $J$- and $H$- bands, respectively.

For MIR data, we used a catalog of \cite{Nayyeri} who provided 3.6 $\micron$ (ch1) and 4.5 $\micron$ (ch2) data taken with Infrared Array Camera \citep[IRAC:][]{Fazio} on board the {\it Spitzer Space Telescope} \citep{Werner}.
This photometric catalog contains 380,858 sources with a 5$\sigma$ detection limit of 6.45 and 3.95 $\mu$Jy in the 3.6 $\micron$ and 4.5 $\micron$ bands, respectively.

Regarding the FIR data, we utilized a recent catalog of 250, 350, and 500 $\micron$ \citep[][C. Pearson in preparation]{Pearson17}.
The data were taken with the Spectral and Photometric Imaging REceiver instrument \citep[SPIRE: ][]{Griffin} on board the {\it Herschel Space Observatory} \citep{Pilbratt}. 
This SPIRE catalog contains 4,820 sources with a 5$\sigma$ detection limit of 9.0, 7.5, and 10.8 mJy at 250, 350, and 500 $\micron$, respectively \citep{Barrufet}.
Note that {\it AKARI} NEP-D was observed by the Photoconductor Array Camera and Spectrometer \citep[PACS: ][]{Poglitsch} at 100 and 160 $\micron$, and the catalog (including 1,384 and 630 sources detected by 100 and 160 $\micron$, respectively) is available \citep{Pearson}. But since the survey footprint is not so large ($\sim$0.7 deg$^{2}$, see Figure \ref{Sky}), we did not use the catalog in this work\footnote{The {\it AKARI} NEP-W was also  partly observed in X-rays with {\it Chandta} \citep{Krumpe}, in the ultraviolet with {\it GALEX} \citep{Buat_17},  at 1.4 GHz with the Westerbork Radio Synthesis Telescope \citep{White}, and at 850 $\micron$ with the Submillimetre Common User Bolometer Array 2 on the James Clerk Maxwell Telescope (H. Shim et al., 2020, in preparation).
But we focus on physical properties of our sample based on the unifirm optical--IR data in this work.}.

Eventually, we used 21 photometric data in total; $g$-, $r$-, $i$-, $z$-, $y$-, $J$-, and $H$-bands and 2.4 (N2), 3.2 (N3), 3.6 (ch1), 4.1 (N4), 4.5 (ch2), 7.0 (S7), 9.0 (S9W), 11.0 (S11), 15.0 (L15), 18.0 (L18W), 24.0 (L24), 250, 350, and 500 $\micron$, even though some of them are often upper limits (e.g., HSC five-bands).

%------------------
% Sample Selection
%------------------
\subsection{Sample Selection}

\begin{figure}
    \centering
    \includegraphics[width=0.45\textwidth]{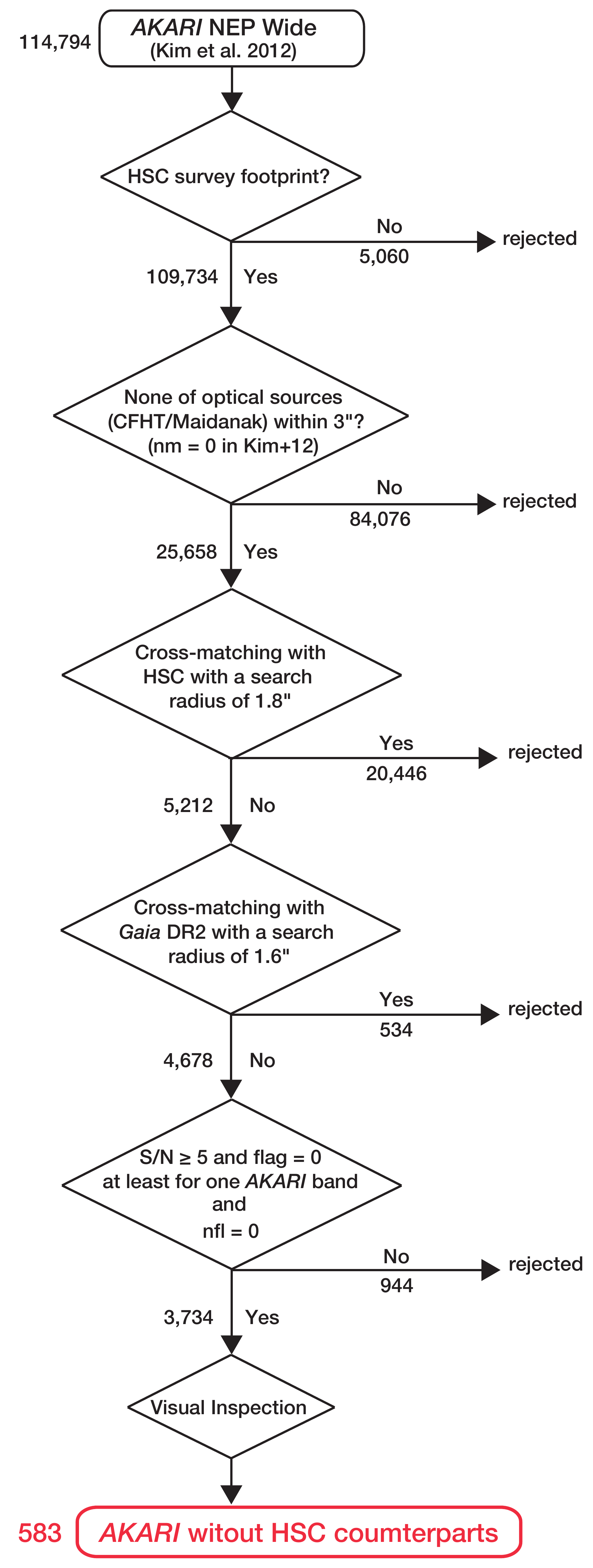}
\caption{Flow chart of the process to select {\it AKARI} sources without HSC counterparts.}
\label{SS}
\end{figure}

The procedure for sample selection is summarized in Figure \ref{SS}.
The candidates of {\it AKARI} sources without HSC counterparts were drawn from the {\it AKARI} NEP-W sample in \cite{Kim} who provided 114,794 sources detected by the IRC.
The 5$\sigma$ detection limit of N2, N3, N4, S7, S9W, S11, L15, L18W, and L24 is 15.42, 13.30, 13.55, 58.61, 67.30, 93.76, 133.1, 120.2, and 274.4 $\mu$Jy, respectively.
Note that we selected candidates and obtained their multi-wavelength information by cross-matching with several catalogs in which coordinates in the {\it AKARI} NEP-W catalog were always referred.
In this work, we attempt a cross-matching with a catalog by using a search radius that is much larger than a typical size of point spread function (PSF) for objects in the catalog.
We then determined a search radius for cross-matching with a catalog as a 3$\sigma$ deviation from mean separation ($\Delta$R.A. and $\Delta$ Decl.) between {\it AKARI} NEP-W and that catalog, in the same manner as S. J. Kim et al. (2020, in preparation).

\begin{figure*}
    \centering
    \includegraphics[width=0.8\textwidth]{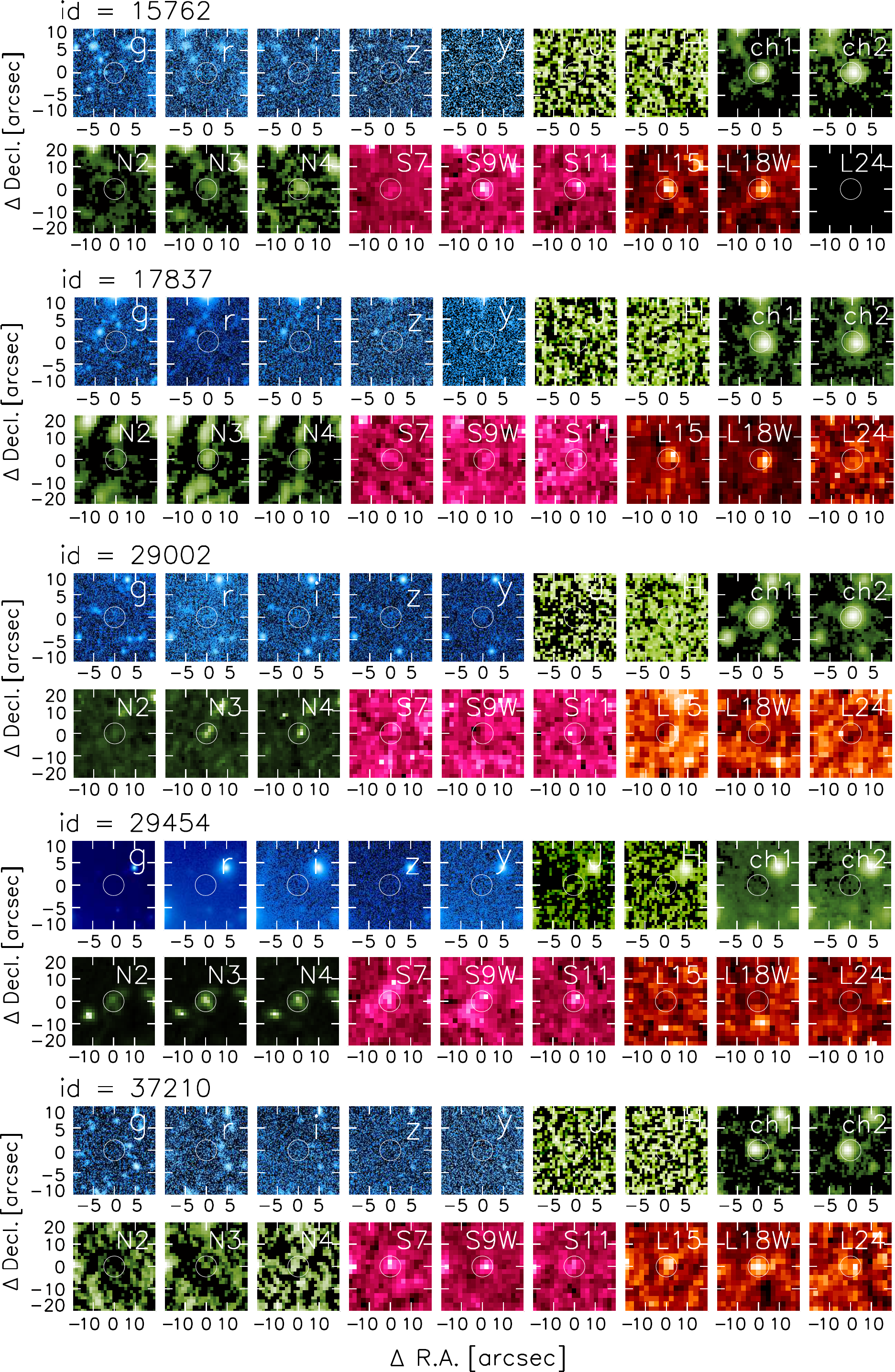}
\caption{Examples of multi-wavelength images ($g$, $r$, $i$, $z$, $y$, $J$, $H$, ch1, ch2, N2, N3, N4, S7, S9W, S11, L15, L18W, and L24, from top left to bottom right) for {\it AKARI} sources without HSC counterparts. R.A. and Decl. are relative coordinate with respect to objects in the {\it AKARI} NEP-W catalog \citep{Kim}. White circles in the images also correspond to the coordinate of {\it AKARI} NEP-W catalog.}
\label{image}
\end{figure*}

We first narrowed the sample to sources within the footprint observed by the HSC.
This is because the HSC data do not cover quite the whole region of the {\it AKARI} NEP-W (see Figure \ref{Sky}), and hence some objects are just unobserved by the HSC.
As a result, 109,734 objects were left out.
We then removed 84,076 objects with {\tt nm\footnote{It provides the number of optical sources matched
to an {\it AKARI} source within 3\arcsec\, \citep[see][]{Kim}.}} $>$ 0 that were reported as having optical counterparts in \cite{Kim}.
They used optical catalogs provided by \cite{HwangN} and \cite{Jeon10} in which the optical data were taken with MegaCam \citep{Boulade} on the Canada-France-Hawaii Telescope (CFHT) and SNUCAM \citep{Im} on the Maidanak observatory, respectively.
Following that, we cross-identified the sample with the HSC catalog \citep{Oi}, which has a sensitivity roughly 5 times deeper than optical catalogs used in \cite{Kim}, providing optically-faint {\it AKARI} sources.
By adopting a search radius of 1.8\arcsec, 20,446 objects were cross-identified.

For $25,658 - 20,446 = 5,212$ {\it AKARI} sources, we removed contaminants.
Because objects with magnitude brighter than $\sim$16 may be saturated in the MegaCam/SNUCAM/HSC images, they were removed from those catalogs before the cross-matching with {\it AKARI} NEP-W catalog.
Therefore, some {\it AKARI} sources in our sample are expected to be bright stars/galaxies.
In order to remove bright objects, we cross-identified the sample with {\it Gaia} DR2 \citep{Gaia,Gaia2} using a search radius of 1.6\arcsec.
The {\it Gaia} DR2 catalog contains point sources with a $g$-band magnitude of 3--16.
As a result, 534 bright objects were removed.\
We then extracted {\it AKARI} objects with 5$\sigma$ detections in at least one {\it AKARI} band and with clean photometry flag\footnote{\cite{Kim} employed {\tt SExtractor} \citep{Bertin} for source detection and photometry \citep[see Section 3 in][for more detail]{Kim}.} (i.e., {\tt flag\_{\rm bandname}\_mag} = 0).
We also applied {\tt nfl} = 0 to select objects unaffected by cosmic rays and/or multiplexer bleed trails in the NIR bands \citep[see Section 2.2. in][]{Kim}, which yielded 3,734 objects.

Finally, we conducted a visual inspection to select reliable candidates in which we supplementarily used multi-wavelength images in $J$- and $H$-bands (FLAMINGOS) and ch1 and ch2 (IRAC) in addition to HSC and {\it AKARI} images.
Consequently, 583 objects were selected as {\it AKARI} sources without HSC counterparts.
Figure \ref{image} shows examples of postage stamp images for our sample.
Why were 3,734 -- 583 = 3,151 objects removed?
One reason is that the edges of the fields of view (FoV) of HSC may be degraded by artifacts.
The {\it AKARI} NEP-W consists of 4-7 exposures by the HSC with a FoV of 1.5\arcdeg \,in diameter through dithering observations.
Therefore, objects around the edges of each exposure frame would be missed from the HSC catalog created by the HSC pipeline \citep{Bosch} although they exist on the HSC image \citep[see ][in details]{Oi}.
Nevertheless, we should note that the visual inspection might be highly dependent on the classifier.
The number of objects selected in this work may have an uncertainty. 
Hence, the population census results such as number counts and volume density of {\it AKARI} sources without HSC counterparts should be addressed in future work, and we focus on an overview of their physical properties in this work.

For those {\it AKARI} sources without HSC counterparts, we compiled a multi-wavelength dataset up to 500 $\micron$.
We cross-identified the sample with FLAMINGOS, IRAC/{\it Spitzer}, and SPIRE/{\it Herschel}.
In our sample, {\it AKARI} coordinates of each source were always used for the cross-matching.
We employed 1.9\arcsec, 2.3\arcsec, and 6.6\arcsec as the search radius for cross-matching with FLAMINGOS, IRAC, and SPIRE, respectively.
As stated, these search radii were determined to be 3$\sigma$ deviations from mean separation between {\it AKARI} objects without HSC counterparts and FLAMINGOS/IRAC/SPIRE catalogs.
Accordingly, 70, 338, and 35 objects were cross-identified with FLAMINGOS, IRAC, and SPIRE, respectively.
We found that 2/70, 4/338 and 2/35 {\it AKARI} objects have two candidate counterparts for FLAMINGOS, IRAC, and SPIRE, respectively.
In this study, we chose the closest object as the counterpart for such cases.
Among {\it AKARI}-bands, 425/583 (73\%), 109/583 (19\%), and 59/583 (10\%) objects are detected in the NIR, MIR-S, and MIR-L, respectively,
We note that 23 objects with L24 detection naturally satisfy DOG criterion, $(r-[24\,\micron])_{\rm AB} > 7.5$ \citep{Dey} owing to the limiting magnitude in the HSC $r$-band (26.7) and {\it AKARI} 24 $\micron$ (17.8), suggesting that our sample selection may preferentially select dusty galaxies (see also Section \ref{S_AV}).

%------------------
% SED fitting with CIGALE
%------------------
\subsection{SED Fitting with {\tt CIGALE}}
\label{s_CIGALE}

\begin{table}
\renewcommand{\thetable}{\arabic{table}}
\centering
\caption{Parameter Ranges Used in the SED Fitting with {\tt CIGALE}} 
\label{Param}
\begin{tabular}{lc}
\tablewidth{0pt}
\hline
\hline
Parameter & Value\\
\hline
\multicolumn2c{Delayed SFH}\\
\hline
$\tau_{\rm main}$ (Myr) 	& 5000.0 \\
$\tau_{\rm burst}$ (Myr)	& 20000\\
$f_{\rm burst}$ 			& 0.00, 0.01\\
age  (Myr) 					& 1000, 5000, 10000\\
\hline
\multicolumn2c{SSP \citep{Bruzual}}\\
\hline
IMF				&	\cite{Salpeter} \\
Metallicity		&	0.02 \\
\hline
\multicolumn2c{Dust Attenuation \citep{CF00}}\\
\hline
$\log$ ($A_{\rm V}^{\rm ISM}$) 		& -2, -1.7, -1.4, -1.1, -0.8, -0.5,\\
									& -0.2 ,0.1, 0.4, 0.7, 1.0 \\
 slope\_ISM & -0.9, -0.7, -0.5 \\
 slope\_BC & -1.3, -1.0, -0.7 \\
\hline
\multicolumn2c{AGN Emission \citep{Fritz}}\\
\hline
$R_{\rm max}/R_{\rm min}$& 60  \\
$\tau_{\rm 9.7}$& 0.3, 0.6 \\
$\beta$& -0.5\\
$\gamma$& 4.0 \\
$\theta$& 100.0\\
$\psi$& 0.001, 60.100, 89.990 \\
$f_{\rm AGN}$& 0.1, 0.3, 0.5, 0.7, 0.9 \\
\hline
\multicolumn2c{Dust Emission \citep{Draine}}\\
\hline
 $q_{\rm PAH}$ & 0.47, 1.77, 2.50, 5.26 \\
 $U_{\rm min}$ & 0.1, 1.0, 10, 50 \\
 $\alpha$ & 1.0, 1.5, 2.0, 2.5, 3.0 \\
 $\gamma$ & 0.01, 0.1, 1.0 \\
 
\hline
\multicolumn2c{Photometric Redshift}\\
\hline
$z_{\rm photo}$	& 0.8, 1.0, 1.5, 2.0, 2.2, 2.4, 2.6, 2.8, 3.0, \\
				& 3.2, 3.4, 3.6, 3.8, 4.0, 4.2\\
\hline
\end{tabular}
\end{table}

We employed {\tt CIGALE} to conduct detailed SED modeling in a self-consistent framework by considering the energy balance between the UV/optical and IR.
This code enables us to handle many parameters, such as star formation history (SFH), single stellar population (SSP), attenuation law, AGN emission, dust emission, and radio synchrotron emission \cite[see e.g.,][]{Boquien14,Buat_14,Buat,Boquien16,Ciesla_17,LoFaro,Toba_19a,Burgarella20,Toba_20a}.
Note that one of the purposes of this work is to compare the physical properties between {\it AKARI} sources with and without HSC counterparts. 
\cite{Tina} constructed {\it AKARI} L18W-selected sample with HSC counterparts, and also conducted the SED fitting with {\tt CIGALE} to derive their physical properties.
Therefore, to compare the resultant quantities under the same conditions, we chose the same models with the same parameter ranges as those \cite{Tina} adopted, except for AGN fraction in the AGN module and $\gamma$ in the dust module (see below).
Parameter ranges used in the SED fitting are tabulated in Table \ref{Param}.

We used a delayed SFH model, assuming a single starburst with an exponential decay \citep{Ciesla_15,Ciesla_16}, 
where we fixed e-folding times of the main stellar population ($\tau_{\rm main}$) and the late starburst population ($\tau_{\rm burst}$), while we parameterized the age of the main stellar population in the galaxy.

We utilized the stellar templates with solar metallicity provided from \cite{Bruzual} assuming the \cite{Salpeter} IMF, and the standard default nebular emission model included in {\tt CIGALE} \citep[see ][]{Inoue}.

Dust attenuation is modeled by using the \cite{CF00} with two different power-law attenuation curves that are parameterized by the power law slope of the attenuation in the interstellar medium (ISM) and birth clouds (BC). 
We also separately parameterized the $V$-band attenuation in the ISM ($A_{\rm V}^{\rm ISM}$).

For AGN emission, we used models provided by \cite{Fritz}.
In order to avoid a degeneracy of AGN templates in the same manner as in \cite{Ciesla_15} and \cite{Toba_19b}, we fixed certain parameters that determine the number density distribution of the dust within the dust torus, i.e., ratio of the maximum to minimum radii of the torus ($R_{\rm max}/R_{\rm min}$), density profile along the radial and the polar distance coordinates parameterized by $\beta$ and $\gamma$ \citep[see equation 3 in][]{Fritz}, and opening angle ($\theta$).
We parameterized the optical depth at 9.7 $\micron$ ($\tau_{\rm 9.7}$) and $\psi$ parameter (an angle between equatorial axis and line of sight) that corresponds to our viewing angle of the torus.
We further parameterized AGN fraction ($f_{\rm AGN}$), that is the contribution of AGN to the total IR luminosity \citep{Ciesla_15}.
Note that we adopted a discrete interval for $f_{\rm AGN}$ compared with \cite{Tina} in which $f_{\rm AGN}$ is a key parameter to investigate the dependences of the fractional AGN contribution on IR luminosity and redshift.
Our sample is basically detected in only several {\it AKARI} bands (and the remaining bands give upper-limits).
This may often not be enough to constrain the $f_{\rm AGN}$ precisely.
Hence we reduced the number of possible values of $f_{\rm AGN}$  considered.

Dust grain emission is modeled by \cite{Draine}.
The model is parameterized by the mass fraction of PAHs ($q_{\rm PAH}$), the minimum radiation field ($U_{\rm min}$), and the power-low slope of the radiation field distribution ($\alpha$) \citep[see Equation 4 in][]{Draine}.
We also parameterized the fraction illuminated with a variable radiation field ranging from $U_{\rm min}$ to $U_{\rm max}$ ($\gamma$) although \cite{Tina} fixed this at $\gamma$ = 1.
We confirmed that parametrizing $\gamma$ gives a better fit to the FIR part of the spectra.

We also parameterized redshift to estimate photometric redshift ($z_{\rm photo}$), because by definition, our sample is optically too faint to obtain spectroscopic redshifts.
Since the optically-dark galaxies tend to be located at $z >2$ \citep[e.g.,][]{Franco,Yamaguchi,Williams,Wang}, we optimized the range of $z_{\rm photo}$, which reduces the computing time.
Aside from being an excellent SED-fitting tool, {\tt CIGALE} is known to be a good estimator of $z_{\rm photo}$, since it utilizes a large number of models covering the whole SED including the MIR-FIR regime.
For example, \cite{Malek} calculated $z_{\rm photo}$ for {\it AKARI} sources in the {\it AKARI} Deep Field South.
They demonstrated that accuracy of the $z_{\rm photo}$ by using the normalized median absolute deviation defined as $\sigma_{\Delta z/(1+z_{\rm spec})}$  = 1.48 $\times$ median($|\Delta z|$/(1+$z_{\rm spec}$)) in the same manner as \cite{Ilbert}.
The resultant $\sigma_{\Delta z/(1+z_{\rm spec})}$ is 0.056, lower than what obtained by a software using mainly optical to NIR data \citep{Ilbert} \citep[see also][]{Barrufet}.
On the other hand, the above {\it AKARI} sources have optical counterparts, i.e., they are moderately bright in the optical, and the sample is limited to the local universe ($z < 0.25$).
Since we constrain the optical SEDs based only on upper limits, we need to investigate the influence of our lack of optical data points on the accuracy of $z_{\rm photo}$ (see Section \ref{ss_photoz}).

In order to ensure the reliability of the derived physical quantities including $z_{\rm photo}$, we extracted {\it AKARI} sources with 5$\sigma$ detections in at least 5 bands among the NIR-FIR, which yields 142/583 objects for SED fitting.
If the signal-to-noise ratio (S/N) at a certain band is greater than 5.0, we used the photometry at that band.
Otherwise, we put 10$\sigma$ upper limits\footnote{{\tt CIGALE} can handle SED fitting of photometric data with upper limits when using the method presented by \cite{Sawicki}. This method computes $\chi^2$ by introducing the error function \citep[see Equations (15) and (16) in][]{Boquien}.} that are drawn from each catalog (see Section \ref{DS}).

%%%%%%%%%%%%%%%%%%%%%%
%    Results
%%%%%%%%%%%%%%%%%%%%%%
\section{Results}
\label{R}

\begin{figure}
    \centering
    \includegraphics[width=0.45\textwidth]{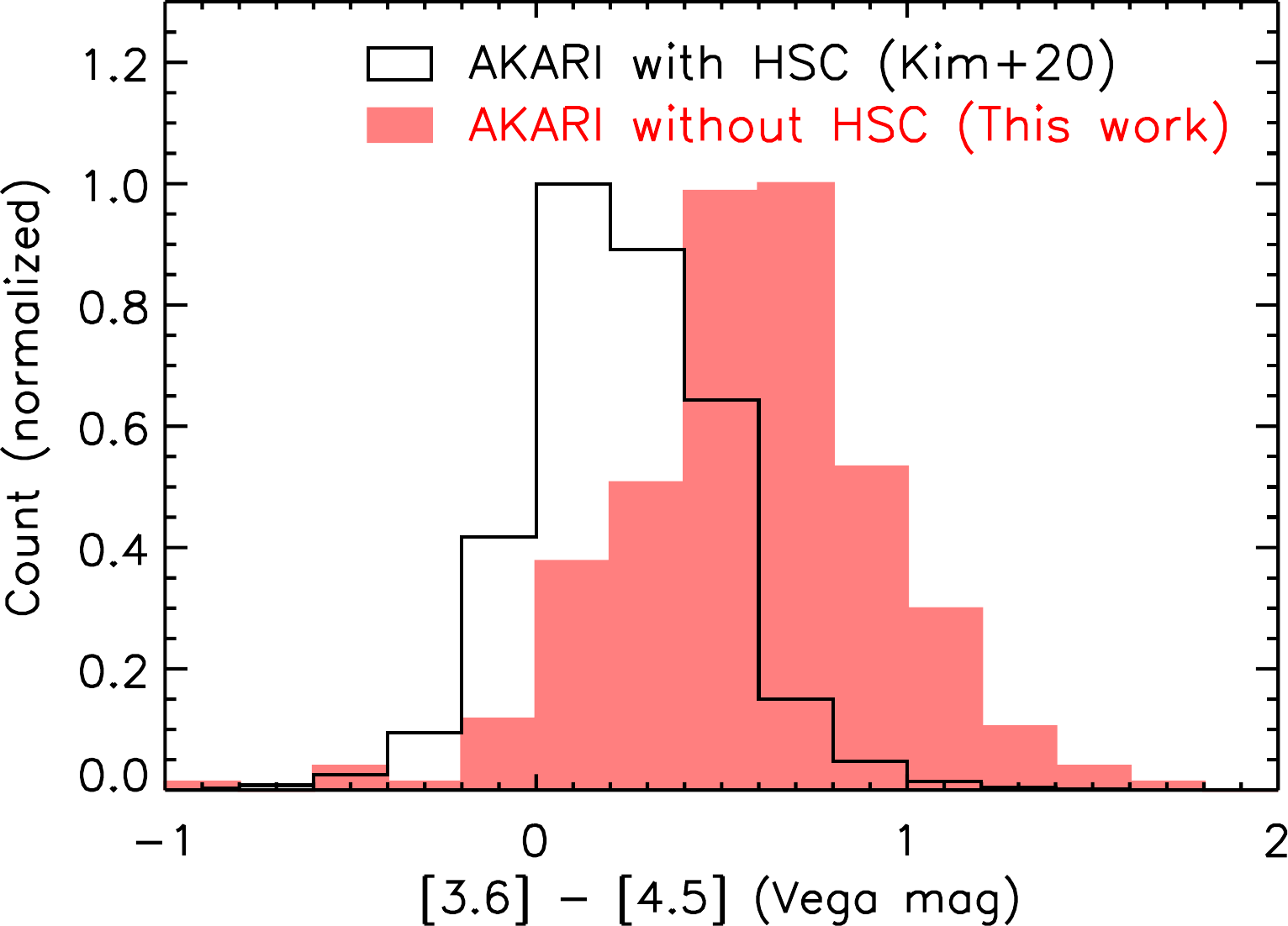}
\caption{Normalized histogram of [3.6]-[4.5] color (in Vega magnitude) of {\it AKARI} sources with (black) and without (red) HSC counterparts.}
\label{Ch12}
\end{figure}

\begin{figure*}
    \centering
    \includegraphics[width=0.9\textwidth]{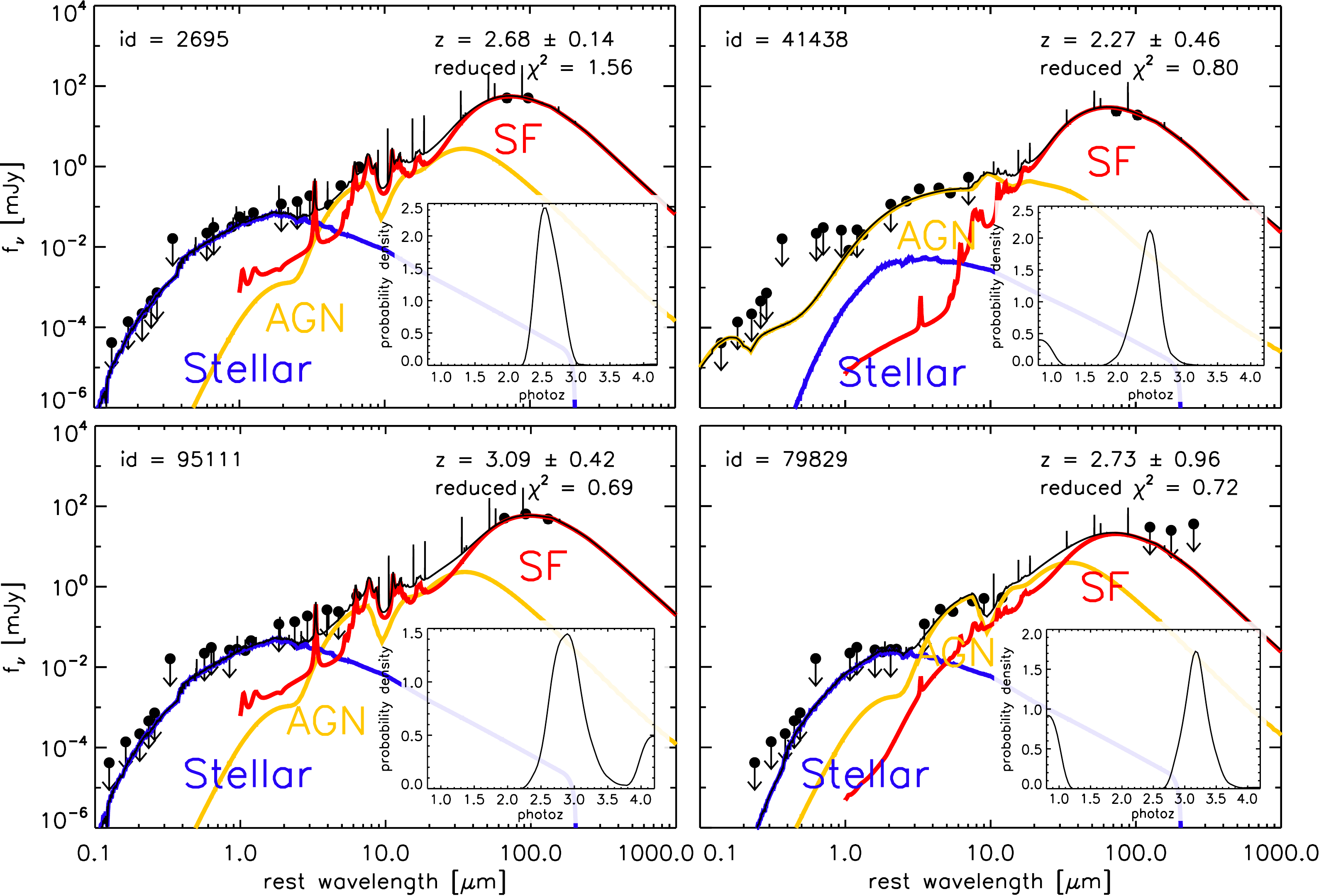}
\caption{Examples of the SED fitting. The black points are photometric data. The contribution from the stellar, AGN, and SF components to the total SED are shown as blue, yellow, and red lines, respectively. The black solid line represents the resultant best-fit SED. The inserted panel shows the probability density distribution of redshift.}
\label{SED}
\end{figure*}

%------------------
% Comparison of IRAC color
%------------------
\subsection{Comparison of IRAC Color}
Before the SED fitting, we check the IRAC ch1 and ch2 color ([3.6] - [4.5] in Vega magnitudes) of our sample objects.
Figure \ref{Ch12} shows IRAC color of 338 {\it AKARI} sources without HSC counterparts.
We also plotted IRAC colors of {\it AKARI} sources with HSC counterparts where we used a multi-wavelength merged {\it AKARI} catalog provided by S. J. Kim et al. (2020, in preparation).
We removed stars with stellarity parameter (that was measured from CFHT $r$-band or Maidanak $R$-band images) greater than 0.8 \citep[see][]{Kim}. This left 42,264 objects for the comparison.

We find that  [3.6]-[4.5] colors of {\it AKARI} sources without HSC counterparts (i.e., our sample) are systematically redder than those of {\it AKARI} sources with HSC counterparts.
\cite{Blecha} reported that W1 (3.4 $\micron$) and W2 (4.6 $\micron$) color (almost same as IRAC ch1 and ch2 color) taken with the the{\it  Wide-field Infrared Survey Explorer} \citep[{\it WISE}:][]{Wright}, is a good indicator of AGN activity and nuclear obscuration.
The AGN luminosity and hydrogen column density peak during the galaxies' coalescence, where W1--W2 color ranges from 0.8 --1.6 (see Figure 1 in \citealt{Blecha}).
This suggests that a fraction of objects in our sample could correspond to luminous, obscured AGN phase.

%------------------
% Result of SED Fitting
%------------------
\subsection{Result of SED Fitting}
\label{S_SED}

Figure \ref{SED} shows examples of the SED fitting with {\tt CIGALE}.
We confirm that 91/142 ($\sim$64\%) objects have reduced $\chi^{2} < 3.0$ while 112/142 ($\sim$79\%) objects have reduced $\chi^{2} < 5.0$.
This means that the data are moderately well fitted with the combination of the stellar, AGN, and SF components by {\tt CIGALE}.
We also confirm that the probability distribution function (PDF) of redshift does not have prominent secondary peaks for $\sim$90\% of our sample in which a peak with 30\% of the primary peak is considered as secondary peak
 (see top panels for objects without showing secondary peak and bottom panels for objects showing secondary peak in the PDF in Figure \ref{SED}).
The typical uncertainty of $z_{\rm photo}$ is about 20\% (see also Section \ref{ss_photoz}).
Hereafter, we will focus on a subsample of 112 {\it AKARI} galaxies with reduced $\chi^{2}$ of their SED fitting smaller than 5.0 in the same manner as \cite{Toba_19b}.

%%%%%%%%%%%%%%%%%%%%%%
%    Discussion
%%%%%%%%%%%%%%%%%%%%%%
\section{Discussions}
\label{D}

\cite{Tina} constructed an 18 $\micron$ (L18W)-selected {\it AKARI} sample with HSC counterparts among which 443 objects have {\it Herschel} detections (SPIRE 250 $\micron$ or PACS 100 $\micron$) and spectroscopic redshifts \citep{Shim,Oi17,HKim,Shogaki}.
In order to derive their physical properties such as AGN luminosity and SFR, they performed SED fitting with {\tt CIGALE}\footnote{We parametrized $\gamma$ in the dust emission model as described in Section \ref{s_CIGALE}. For a fair comparison, we re-performed the SED fitting with parametrization of $\gamma$ for the sample provided by \cite{Tina}.}.
What is the difference in the physical properties between {\it AKARI} objects with and without HSC counterparts?
To ensure the relatability of the SED fitting, we will use 317 sources with reduced $\chi^{2} < 5.0$ in sample provided by \cite{Tina} for the following discussion.

%------------------
%   stellar ,mass
%------------------
\subsection{Stellar Mass}
\label{S_M}

\begin{figure}
    \centering
    \includegraphics[width=0.45\textwidth]{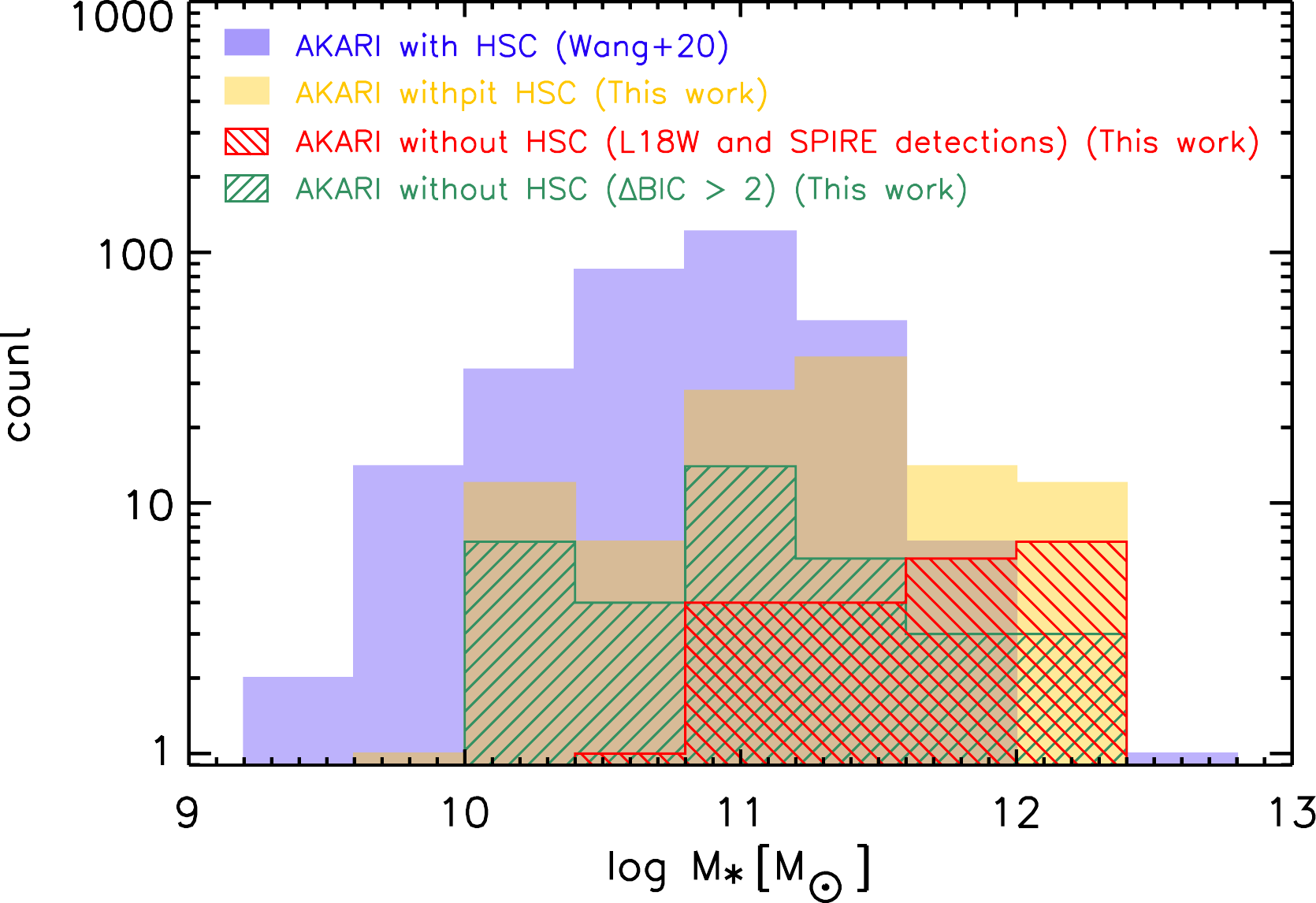}
\caption{Histogram of the stellar mass of {\it AKARI} sources with (blue) and without (yellow) HSC counterparts. The red line represents stellar mass of our sample with L18W and SPIRE detections. Green line represents our sample with $\Delta$BIC $>$ 2.0 (see Section \ref{s_BIC}).}
\label{M}
\end{figure}

Figure \ref{M} shows the stellar mass for {\it AKARI} sources with and without HSC counterparts.
We find that the stellar mass of our sample galaxies is systematically larger than that of {\it AKARI} sources with HSC counterparts \citep{Tina}.
The average stellar mass of {\it AKARI} sources with and without HSC counterparts is $\log\, (M_{*}/M_{\sun})$ = 10.8 and 11.3, respectively.
A two-sided Kolmogorov-Smirnov (K-S) test rules out a hypothesis that two samples are drawn from the same distribution at $ >$99.9\% significance.

One caution here is that the procedure of sample selection in this work differs from  \cite{Tina}; AKARI--HSC objects in \cite{Tina} requires both L18W and Herschel detections, while our sample is not necessarily detected by them.
Therefore, we extracted 22 objects that satisfy the sample selection criteria in \cite{Tina}.
Their average $\log\, (M_{*}/M_{\sun})$ is 11.7, also significantly larger than sample in \cite{Tina}, which is also confirmed by the K-S test with $ >$99.9\% significance.

Figure \ref{M} also shows objects with the Bayesian information criterion \citep[BIC; ][]{Schwarz} greater than 2, which indicates that those objects need the AGN component to give a better SED fitting (see Section \ref{s_BIC} for more detail). 
We confirmed that there is no systematic difference in those objects and others.

%------------------
%     AV
%------------------
\subsection{Dust Attenuation in the ISM}
\label{S_AV}

\begin{figure}
    \centering
    \includegraphics[width=0.45\textwidth]{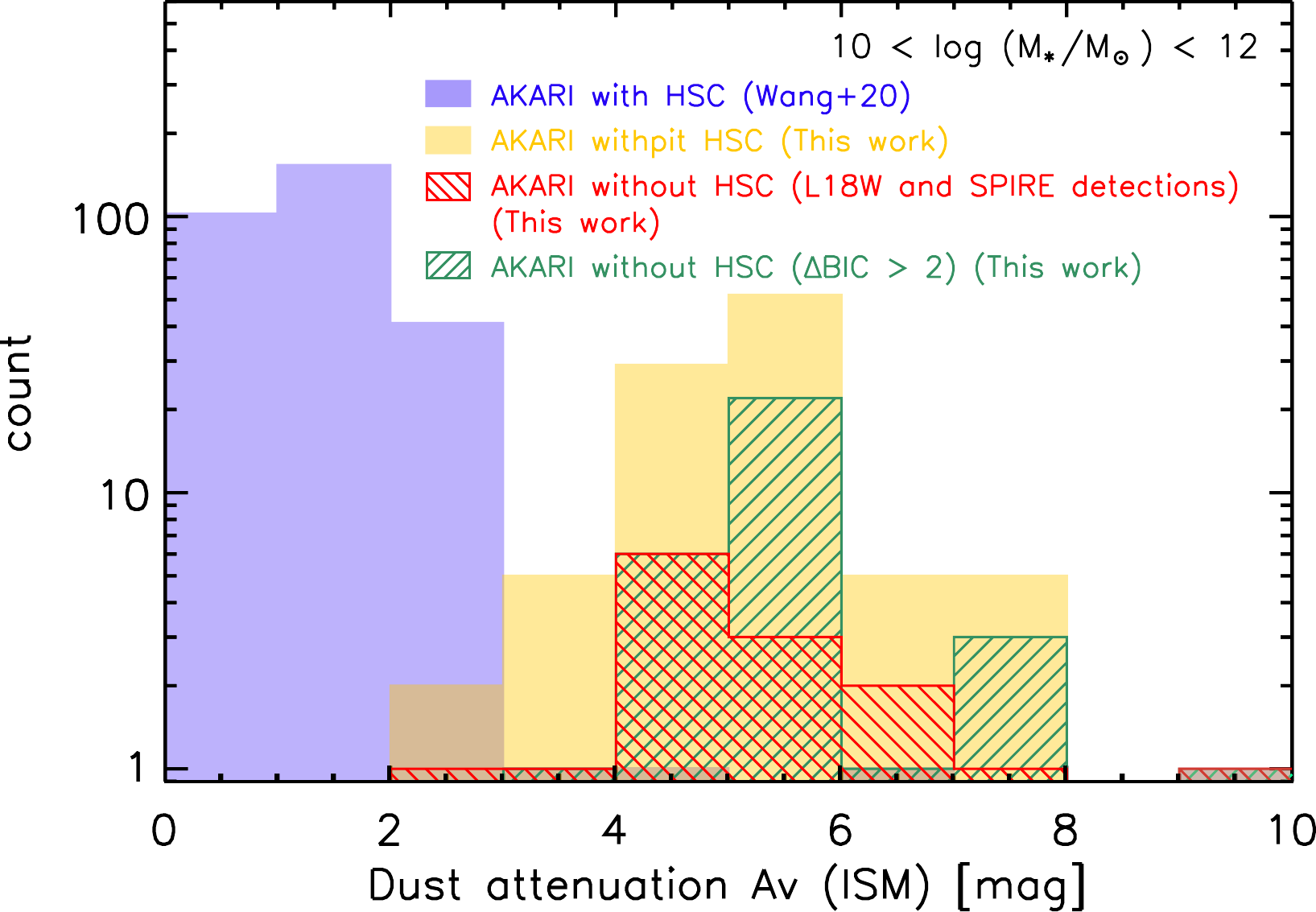}
\caption{Histogram of the $V$-band attenuations in the ISM of {\it AKARI} sources with (blue) and without (yellow) HSC counterparts. The red line represents $A_{\rm V}^{\rm ISM}$ of our sample with L18W and SPIRE detections.  Green line represents our sample with $\Delta$BIC $>$ 2.0. The objects with $10 < \log\, (M_{*}/M_{\sun}) < 12$ are plotted in each sample.}
\label{AV}
\end{figure}

We then compare the $V$-band attenuation in the ISM ($A_{\rm V}^{\rm ISM}$) for {\it AKARI} sources with and without HSC counterparts.
Before the comparison, we should note that the dust attenuation may depend on the stellar mass \citep[e.g.,][]{Buat_12,Buat}, and the stellar mass distributions in the two samples are different as discussed in Section \ref{S_M}.
Therefore, we extracted objects in an overlapped stellar mass range, i.e., $10 < \log\, (M_{*}/M_{\sun}) < 12$ in the {\it AKARI} sample with/without HSC counterparts (see Figure \ref{M}).

Figure \ref{AV} shows the $V$-band attenuation in the ISM ($A_{\rm V}^{\rm ISM}$) for {\it AKARI} sources with and without HSC counterparts.
These objects are stellar-mass matched samples.
The average $A_{\rm V}^{\rm ISM}$ of {\it AKARI} sources with and without HSC counterparts is 1.26 and 5.16 mag, respectively.
We find that the attenuation of our sample is systematically much larger than that of {\it AKARI} sources with HSC counterparts \citep{Tina}, which is supported by the K-S test with $ >$99.9\% significance.
This result seems to be reasonable, given a fact that our sample is undetected even by the HSC, and its optical light is expected to be suppressed heavily by enshrouding dust.

One caution is that the redshift distributions of {\it AKARI} objects in \cite{Tina} and this study are different (see Section \ref{zLAGN}).
Given an overlapped redshift range ($0.8 < z < 2.0$) between two samples (see Figure \ref{LAGN}), the average $A_{\rm V}^{\rm ISM}$ of sample in \cite{Tina} and this work is 1.77 and 5.23 mag, respectively.
This suggests that our sample is intrinsically affected by dust extinction compared with {\it AKARI} with HSC counterparts.

%------------------
%    z vs. LIR
%------------------
\subsection{AGN Luminosity as a Function of Redshift}
\label{zLAGN}

\begin{figure}
    \centering
    \includegraphics[width=0.45\textwidth]{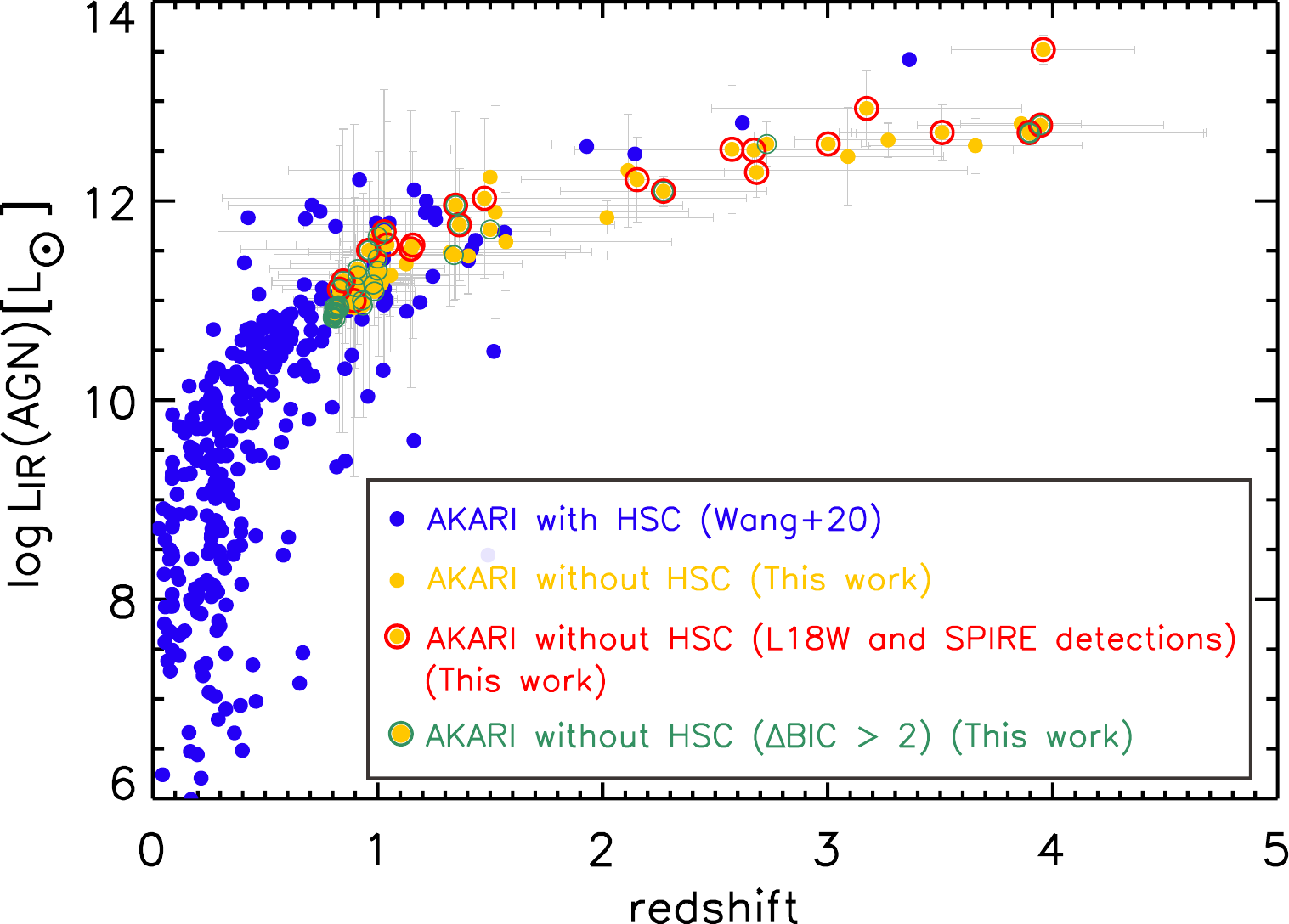}
\caption{IR luminosity contributed from AGN as a function of redshift. Blue and yellow points represent {\it AKARI} sources with and without HSC counterparts, respectively. Yellow and red circles represent our sample with L18W and SPIRE detections. Yellow and green circles represent our sample with $\Delta$BIC $>$ 2.0.}
\label{LAGN}
\end{figure}

Figure \ref{LAGN} shows the IR luminosity contributed from AGN as a function of redshift.
We find that {\it AKARI} objects without HSC counterparts tend to be located in higher redshifts compared to those with HSC counterparts: the mean redshift of {\it AKARI} objects with and without HSC counterparts is $z\sim$ 0.46 and 1.34, respectively.
We also find that objects with L18W and SPIRE detections tend to have large AGN luminosity compared with others in our sample.

Note that the parent sample for these objects is the same, i.e., a flux-limited sample drawn from the {\it AKARI} NEP-W catalog \citep{Kim}.
Thus {\it AKARI} sources are expected to lie in the same sequence on the redshift--luminosity plane, regardless of HSC detection.
We confirm that samples in \cite{Tina} and this work are continuously distributed in that plane, indicating that {\tt CIGALE} securely estimated redshift and luminosity.
Indeed, given the overlapped redshift range of $0.8 < z < 2.0$, the mean AGN luminosity of our sample is $\log (L_{\rm IR} {\rm (AGN)}/L_{\sun}) \sim$ 11.2, is comparable to the sample in \cite{Tina}. 

The total IR luminosity of {\it AKARI} sources without HSC counterparts is also larger than those with HSC counterparts, with a mean $\log\,(L_{\rm IR}/L_{\sun})$ of 11.4 and 12.2, respectively.
Actually, about 65\% and 15\% of our sample is ultra-luminous IR galaxies \citep[ULIRGs:][]{Sanders} and hyper-luminous IR galaxies \citep[HyLIRGs:][]{Rowan-Robinson} with $L_{\rm IR}$ greater than $10^{12}$ and $10^{13}$ $L_{\sun}$, respectively.
This is in good agreement with a fact that dusty galaxies with extreme optical--IR color tend to be luminous in the IR \citep[e.g.,][]{Tsai,Toba_16,Toba_18,Fan,Toba_20b}.

%------------------
%  AGN fraction
%------------------
\subsection{AGN fraction}
\label{s_fAGN}
\begin{figure}
    \centering
    \includegraphics[width=0.45\textwidth]{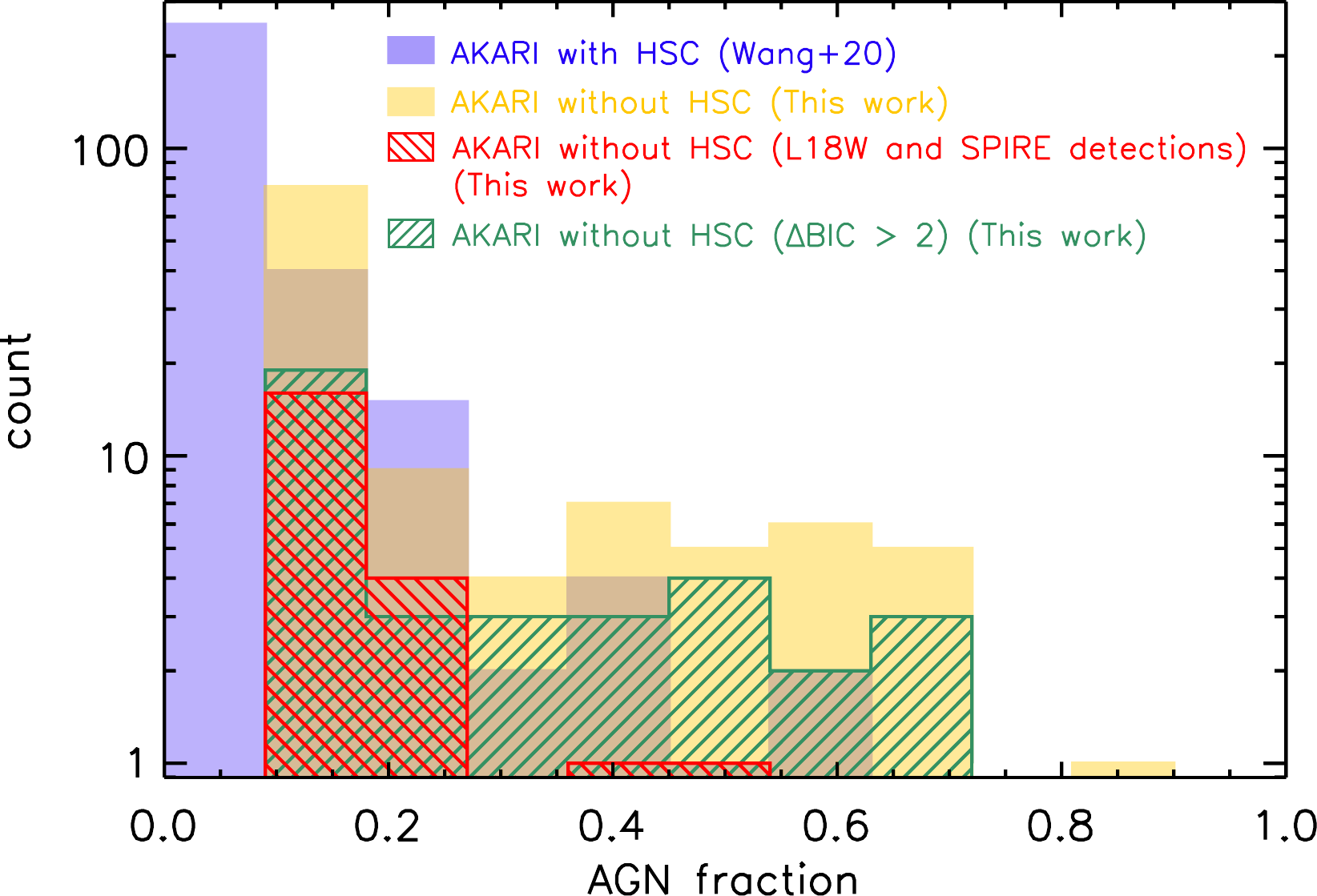}
\caption{Histograms of AGN fraction for {\it AKARI} sources with (blue) and without (yellow) HSC counterparts. The red line represents $f_{\rm AGN}$ of our sample with L18W and SPIRE detections. Green line  represent $f_{\rm AGN}$ of our sample with BIC $>$ 2.0 (see Section \ref{s_BIC}).}
\label{fAGN}
\end{figure}

Figure \ref{fAGN} shows the AGN fraction ($f_{\rm AGN}$),  $L_{\rm IR}$ (AGN) / $L_{\rm IR}$. 
The mean AGN fraction of {\it AKARI} sources with and without HSC counterparts is 0.06 and 0.22, respectively, indicating our sample tends to harbor AGNs.
We also find that $f_{\rm AGN}$ of objects with L18W and SPIRE detection is 0.16, smaller than average of all objects in our sample.
On the other hand, given the overlapped redshift range ($0.8 < z < 2.0$), the mean $f_{\rm AGN}$ of {\it AKARI} sources with and without HSC counterparts is 0.12 and 0.23, respectively.
This could indicate that one of the reasons for the difference in $f_{\rm AGN}$ between the two samples may be the difference in sample selection and redshift.
Other possible uncertainties caused by a limited number of MIR data are discussed in Section \ref{s_BIC}.

%------------------
%      SFR
%------------------
\subsection{Star Formation Rate}

\begin{figure}
    \centering
    \includegraphics[width=0.45\textwidth]{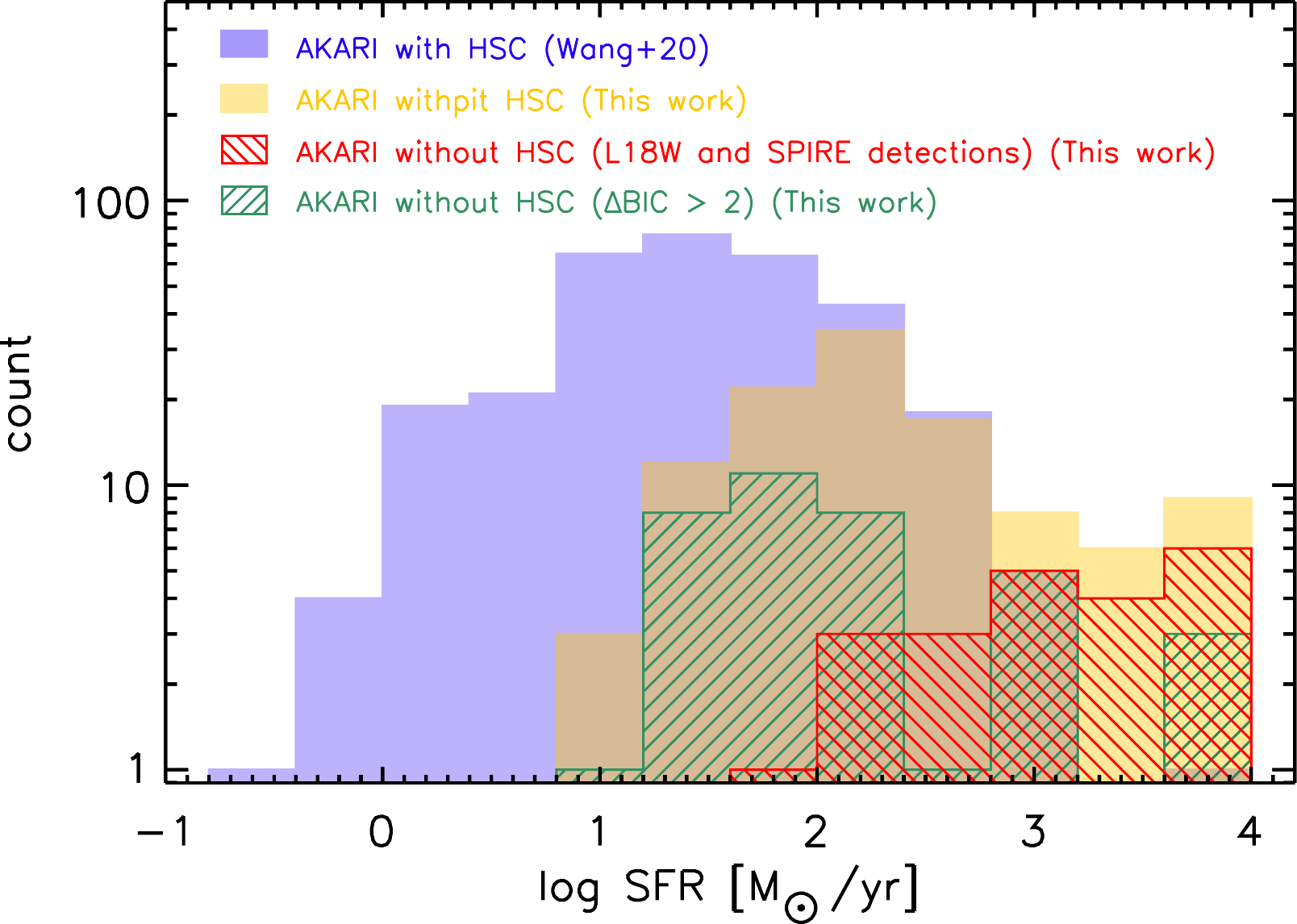}
\caption{Histograms of SFR for {\it AKARI} sources with (blue) and without (yellow) HSC counterparts. The red and green lines represent the SFR of our sample with L18W and SPIRE detections, and with $\Delta$BIC $>$ 2.0, respectively.}
\label{SFR}
\end{figure}

Finally, we compare the SFR in two samples as shown in Figure \ref{SFR}.
The SFR is converted from dust luminosity using a relation provided by \cite{Kennicutt} (see also \citealt{Hirashita}).

We find that the SFR of our sample is systematically higher than that of {\it AKARI} with HSC objects.
The average SFR of {\it AKARI} objects with and without HSC counterparts is about 91 and 758 $M_{\sun}$ yr$^{-1}$, respectively.
If we focus on our sample with L18W and SPIRE detections, the mean SFR is 2240 $M_{\sun}$ yr$^{-1}$ because of requirement of FIR detection.
These results are consistent with previous works reporting that SFRs of dusty galaxies tend to be high \citep[e.g.,][]{Ikarashi,Toba_17b,Yamaguchi,Fan}.
On the other hand, if we compare SFRs of sample in this work and \cite{Tina} in an overlapped redshift range ($0.8 < z < 2.0$), the mean SFR of {\it AKARI} with and without HSC counterparts is 319 and 207 $M_{\sun}$ yr$^{-1}$, respectively.
This indicates that the observed difference in SFR may be due to the redshift difference.

%------------------
%  Reliability of the SED modeling
%------------------
\subsection{Reliability of the SED Analysis}
\label{Rel}
\subsubsection{Influence of Dataset without Optical Photometry on the SED-based Photometric Redshift}
\label{ss_photoz}

We showed that our estimate of $z_{\rm photo}$ is expected to be secure in some ways (see Sections \ref{s_CIGALE} and \ref{zLAGN}), but these arguments are indirect.
Although \cite{Malek} demonstrated the accuracy of $z_{\rm photo}$ based on {\tt CIGALE} for an {\it AKARI} sample at $z_{\rm spec} < 0.25$, they utilized optical data points for $z_{\rm photo}$ estimation, and the redshift range for the sample differs from our sample.
We need to investigate how the lack of optical data affects the accuracy of $z_{\rm photo}$ for {\it AKARI} objects at $z > 0.8$. 
Hence, we estimate $z_{\rm photo}$ of {\it AKARI} objects with $z_{\rm spec}$ in the band-merged catalog (S. J. Kim et al. 2020, in preparation).
In order to calculate the $z_{\rm photo}$ under the same condition as we performed so far, we extracted objects with $z_{\rm spec} > 0.8$ and with 5-band detections in NIR-FIR, yielding 100 objects.
Note that although the sample is detected by the HSC, we ``artificially'' multiplied the HSC flux densities by 10 and treated them as 10$\sigma$ upper limits.
We then executed the SED fitting with {\tt CIGALE} in the exactly same manner as what we described in Section \ref{s_CIGALE}.

\begin{figure}
    \centering
    \includegraphics[width=0.45\textwidth]{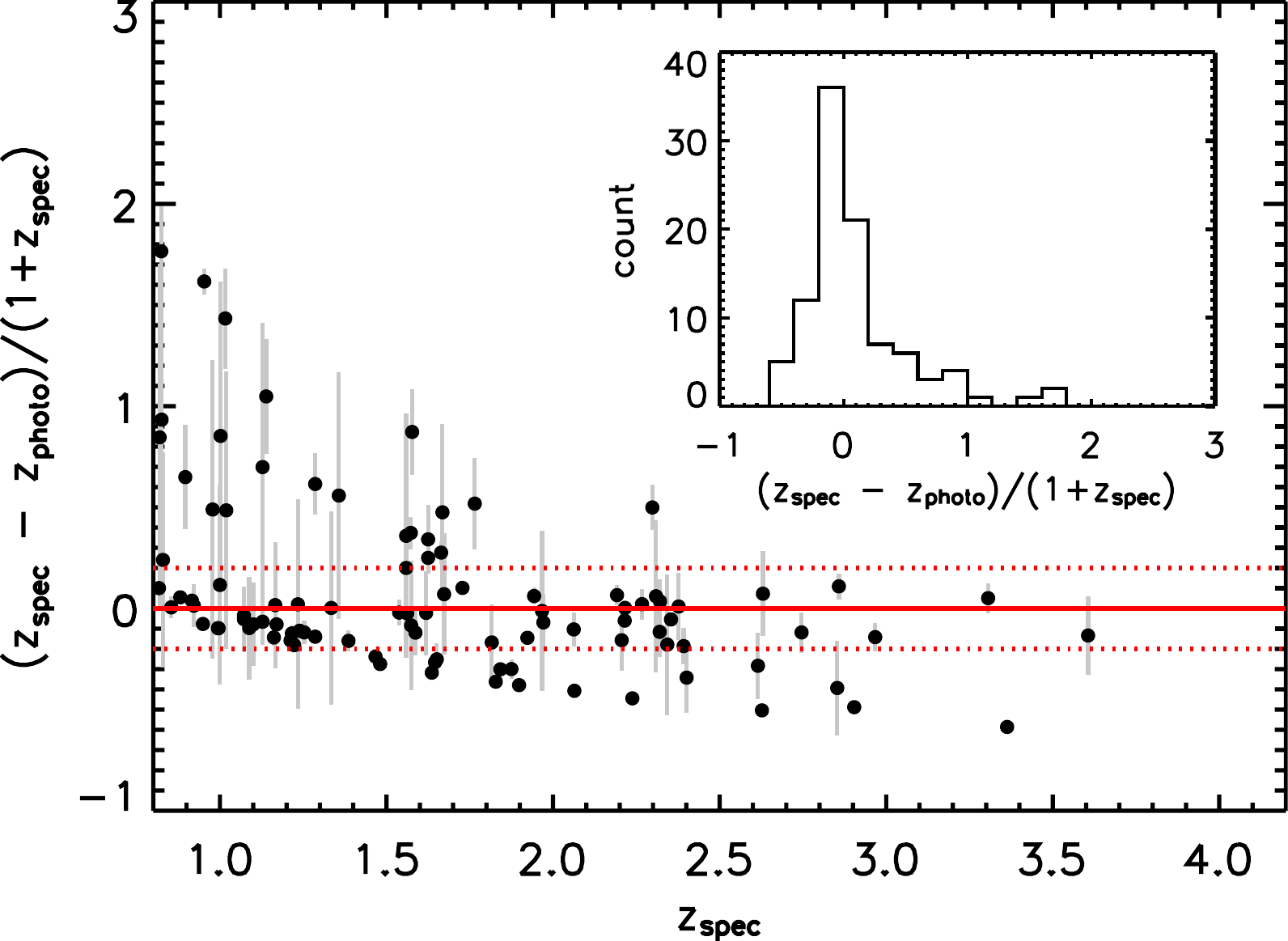}
\caption{Relative difference between $z_{\rm photo}$ and $z_{\rm spec}$, $(z_{\rm spec}-z_{\rm photo})/(1+z_{\rm spec})$ as a function of $z_{\rm spec}$.The red solid line represents $z_{\rm photo} - z_{\rm spec} = 0$ and the dotted lines represent $(z_{\rm spec}-z_{\rm photo})/(1+z_{\rm spec}) = \pm 0.2$. The inserted panel shows the histogram of  $(z_{\rm spec}-z_{\rm photo})/(1+z_{\rm spec})$.} 
\label{spec_photo}
\end{figure}

Figure \ref{spec_photo} shows the relative difference between $z_{\rm photo}$ and $z_{\rm spec}$, i.e., $(z_{\rm spec}-z_{\rm photo})/(1+z_{\rm spec})$ as a function of $z_{\rm spec}$.
The mean and standard deviation of $(z_{\rm spec}-z_{\rm photo})/(1+z_{\rm spec})$ is $0.09 \pm 0.45$, and the accuracy of $z_{\rm photo}$, $\sigma_{\Delta z/(1+z_{\rm spec})}$ is 0.23, which are worse than those reported by \cite{Malek}.
We find that $\sim 56$\% of objects have $|(z_{\rm spec}-z_{\rm photo})|/(1+z_{\rm spec})$ $< 0.2$. 
This may be partly because the AKARI--HSC objects at $z > 0.8$ are biased toward optically-bright type 1 AGNs, for which $z_{\rm photo}$ is harder to estimate precisely, compared with other galaxy types.
Actually,  $\sim 90$\% of objects plotted in Figure \ref{spec_photo} are spectroscopically confirmed type 1 AGNs, which may induce a large deviation of $(z_{\rm spec}-z_{\rm photo})/(1+z_{\rm spec})$.

%Bayesian Information Criterion}
\subsubsection{Bayesian Information Criterion}
\label{s_BIC}
An another potential issue raised by the SED fitting with a limited number of data points in the MIR may be an uncertainty of AGN contribution to the total SED, although we used upper-limits at a MIR band if an object is not detected at that band.
To test the requirement to add an AGN component to the SED fitting, we compute the Bayesian information criterion (BIC) for two fits that are derived with and without AGN component.
The BIC is defined as BIC = $\chi^{2}$ + $k$ $\times$ ln($n$), where $\chi^{2}$ is non-reduced chi-square, $k$ is the number of degrees of freedom (DOF), and $n$ is the number of photometric data points used for the fitting, respectively.
We then compare the results of two SED fittings without/with AGN module by using $\Delta$BIC = BIC$_{\rm woAGN}$ -- BIC$_{\rm wAGN}$.
The resultant $\Delta$BIC tells whether the AGN model is required to give a better fit with taking into account the difference in DOF \citep[e.g.,][see also \citealt{Aufort}]{Ciesla_18,Buat_19}.

\begin{figure}
    \centering
    \includegraphics[width=0.45\textwidth]{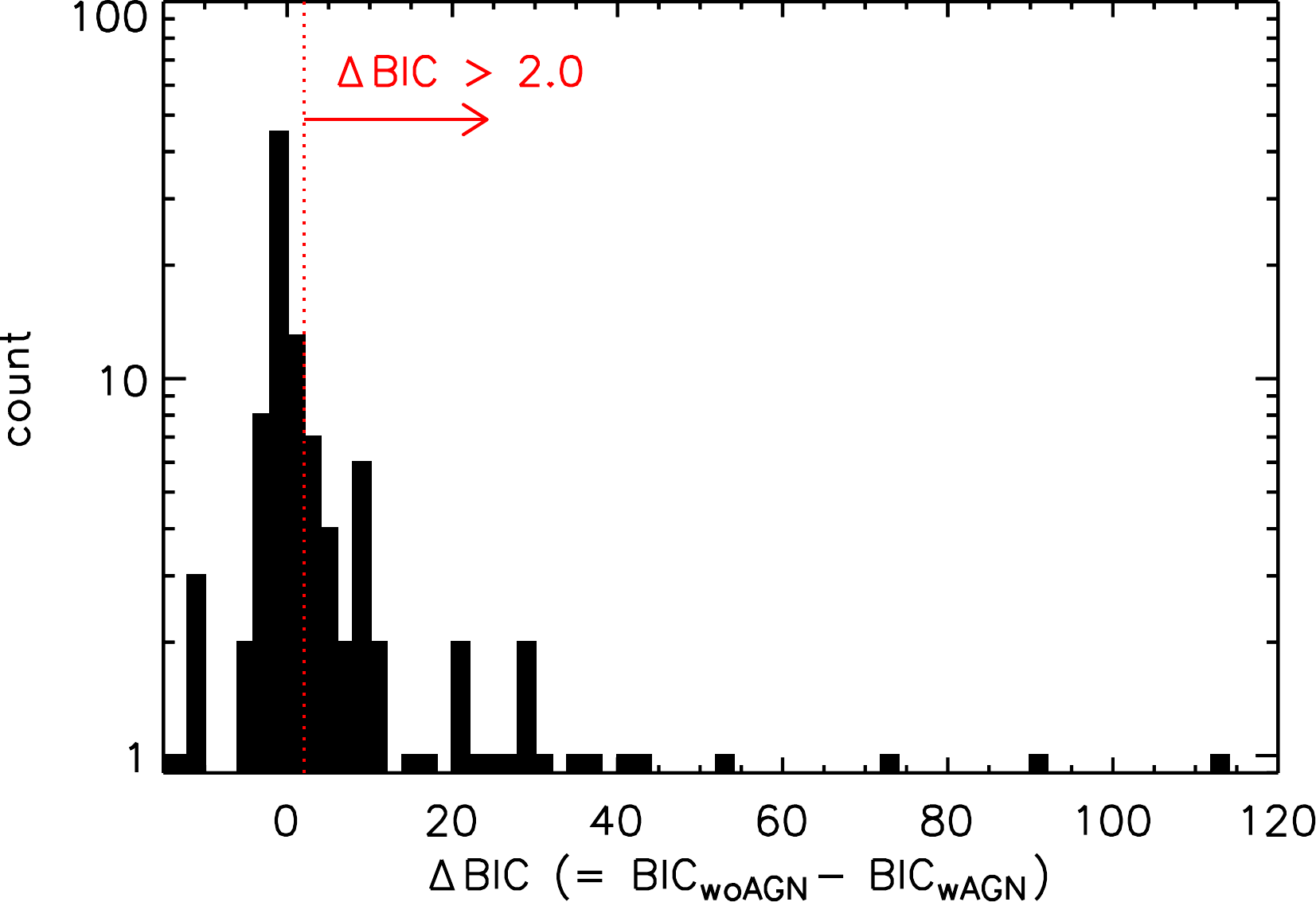}
\caption{Histogram of $\Delta$BIC = BIC$_{\rm woAGN}$--BIC$_{\rm wAGN}$ for {\it AKARI} sources without HSC counterparts. The red dotted line corresponds to $\Delta$BIC = 2.}
\label{BIC}
\end{figure}

Figure \ref{BIC} shows the histogram of $\Delta$BIC for {\it AKARI} sources without HSC counterparts.
If $\Delta$BIC is larger than two, this indicates that adding the AGN component provides a better fit than without adding it \citep{Liddle,Stanley} (see also \citealt{Ciesla_18,Buat_19} who set a higher threshold for $\Delta$BIC).
Otherwise, there is no significant difference between two fits with/without AGN model.
We find that only 31\%  of object fits satisfy $\Delta$BIC $> 2$. This suggests it may be difficult to constrain the AGN activity for many cases, and thus AGN fraction may have a large uncertainty given a limited number of photometric detections in the MIR regime.
On the other hand, Figures \ref{M}--\ref{SFR} also show objects with $\Delta$BIC $> 2$.
We find that there is no systematic difference between objects with  $\Delta$BIC $> 2$ and others, especially for AGN fraction (see Figure \ref{fAGN}), which suggests that overall trends discussed in Section \ref{s_fAGN} may not be changed even if taking into account the BIC.

%Mock Analysis
\subsubsection{Mock Analysis}
Finally, we check whether or not the derived physical properties in this work can actually be estimated reliably, given the uncertainty of the photometry.
We conduct a mock analysis that is a procedure provided by {\tt CIGALE} \citep[see e.g.,][for more detail]{Buat_12,Buat_14,Ciesla_15,LoFaro,Boquien,Toba_19b}.

\begin{figure}
    \centering
    \includegraphics[width=0.45\textwidth]{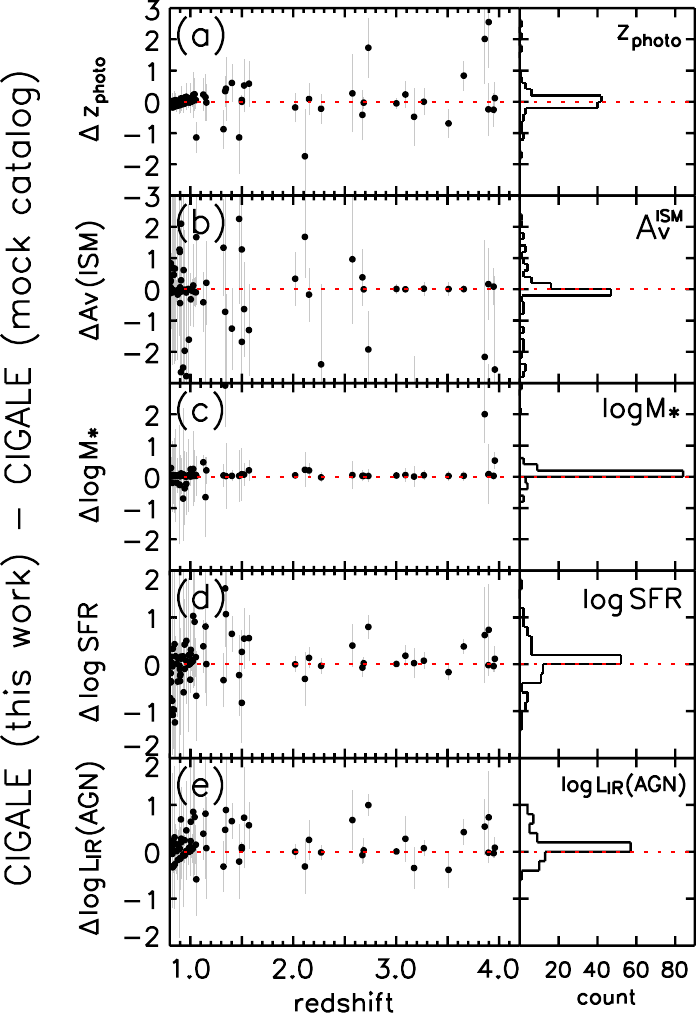}
\caption{The differences in $z_{\rm photo}$, $V$-band attenuation in the ISM, stellar mass, SFR, and $L_{\rm IR}$(AGN) derived from CIGALE in this work and those derived from the mock catalog. (a) $\Delta z_{\rm photo}$, (b) $\Delta A_{\rm V}^{\rm ISM}$, (c) $\Delta \log \, M_{*}$, (d) $\Delta \log{\rm SFR}$, and (e) $\Delta \log L_{\rm IR}$(AGN) as a function of redshift. The right panels show a histogram of each quantity. The red dotted lines are the $\Delta$= 0.}
\label{mock}
\end{figure}

Figure \ref{mock} shows the differences in $z_{\rm photo}$, $V$-band attenuation in the ISM ($A_{\rm V}$), stellar mass, SFR, and $L_{\rm IR}$(AGN) derived from {\tt CIGALE} in this work and those derived from the mock catalog as a function of redshift. 
The mean and standerd deviations of $\Delta z_{\rm photo}$, $\Delta A_{\rm V}^{\rm ISM}$, $\Delta \log \, M_{*}$, $\Delta \log{\rm SFR}$, and $\Delta \log L_{\rm IR}$(AGN) are $\Delta z_{\rm photo} = 0.03 \pm 0.47$, $\Delta A_{\rm V}^{\rm ISM} = -0.23 \pm 1.17$, $\Delta \log \, M_{*} = 0.08 \pm 0.37$, $\Delta \log{\rm SFR} = 0.03 \pm 0.43$, and  $\Delta \log L_{\rm IR}$(AGN) = 0.11 $\pm$ 0.30,  respectively. 
In particular, we can see a large deviation for objects at $z < 1.5$.
This trend can also be seen in Figure \ref{spec_photo}.
These results indicate that any physical quantities derived by the SED fitting especially for objects at $z < 1.5$ may be significantly affected by the limited number of photometric data points.
On the other hand, we could ensure reliable physical quantities for our sample at $z > 1.5$.

%%%%%%%%%%%%%%%%%%%%%%
%     Summary
%%%%%%%%%%%%%%%%%%%%%%
\section{Summary}
\label{Sum}

In this paper, we report the physical properties of {\it AKARI} sources that do not have optical counterparts in the HSC/Subaru catalog \citep{Oi}.
The parent sample is drawn from IR sources in the {\it AKARI} NEP-W field \citep{Kim}.
By using {\it AKARI}, HSC, {\it Gaia}, FLAMINGOS/KPNO and IRAC/{\it Spitzer} catalogs and images, we select 583 objects as optically-dark IR sources without HSC counterparts.
Thanks to the continuous filters of {\it AKARI} in the MIR and multi-wavelength data up to the FIR (SPIRE/{\it Herschel}), we successfully pin down their optical--FIR SEDs, even if flux densities in some bands are upper limits.
We compare the physical properties derived by the {\tt CIGALE} SED fitting between {\it AKARI} objects without HSC counterparts and mid-IR selected {\it AKARI} objects with HSC counterparts \citep{Tina}.
We find that {\it AKARI} sources without HSC counterparts have systematically redder 3.6 $-$ 4.5 $\micron$ color compared with {\it AKARI} sources with HSC counterparts.
With all the caveats discussed in Section \ref{Rel} in mind, we find that our sample tends to be located at high redshifts up to $z\sim 4$, and has larger AGN luminosity, SFR, and $V$-band dust attenuation in the ISM, compared with {\it AKARI} sources with HSC counterparts.
Although this is partly due to the Malmquist bias, these results indicate that {\it AKARI} objects without HSC counterparts are heavily dust-obscured SFGs/AGNs at $z\sim$ 1-4 that may be missed by the previous optical surveys.

The launch of the {\it James Webb Space Telescope} \citep[{\it JWST}:][]{Gardner} is approaching, and {\it AKARI} NEP should be an attractive field for {\it JWST} \citep[e.g.,][]{Jansen}.
Since the {\it AKARI} NEP has multi-wavelength data from X-ray to radio, in which the filter sets of {\it AKARI} are similar to those of {\it JWST}, this work establishes a benchmark for forthcoming dusty SFGs/AGNs studies with {\it JWST} and provides specially interesting optically dark IR sources for {\it JWST} study.

\acknowledgments
We gratefully acknowledge the anonymous referee for a careful reading of the manuscript and very helpful comments.

%AKARI
This research is based on observations with AKARI, a JAXA project with the participation of ESA.

%Gaua https://gea.esac.esa.int/archive/documentation/GDR2/Miscellaneous/sec_credit_and_citation_instructions/
This work has made use of data from the European Space Agency (ESA) mission {\it Gaia} (\url{https://www.cosmos.esa.int/gaia}), processed by the {\it Gaia} Data Processing and Analysis Consortium (DPAC, \url{https://www.cosmos.esa.int/web/gaia/dpac/consortium}). Funding for the DPAC has been provided by national institutions, in particular the institutions participating in the {\it Gaia} Multilateral Agreement.

%Spitzer
This work is based on observations made with the Spitzer Space Telescope, which is operated by the Jet Propulsion Laboratory, California Institute of Technology under a contract with NASA. Support for this work was provided by NASA through an award issued by JPL/Caltech

%Herschel
Herschel is an ESA space observatory with science instruments provided by European-led Principal Investigator consortia and with important participation from NASA.

%Ichiro
Numerical computations/simulations were carried out (in part) using the SuMIRe cluster operated by the Extragalactic
OIR group at ASIAA.

%KAKENHI
This work is supported by JSPS KAKENHI Grant numbers 18J01050 and 19K14759 (Y.Toba), JP18J40088 (R.Momose), and 17K05384 (Y.Ueda).
Y.Toba and T.Goto acknowledge the support by the Ministry of Science and Technology of Taiwan, MOST 108-2112-M-001-014- and 108-2628-M-007-004-MY3.
R.Momose acknowledges a Japan Society for the Promotion of Science (JSPS) Fellowship at Japan. T.Hashimoto is supported by the Centre for Informatics and Computation in Astronomy (CICA) at National Tsing Hua University (NTHU) through a grant from the Ministry of Education of the Republic of China (Taiwan). T.Miyaji is supported by CONACyT 252531 and UNAM-DGAPA (PASPA and PAPIIT IN111319).

\vspace{5mm}
\facilities{{\it AKARI}, Subaru, KPNO:2.1m, {\it Spitzer}, {\it Herschel}.}


\software{IDL, IDL Astronomy User's Library \citep{Landsman}, {\sf CIGALE} \citep{Boquien}, {\tt TOPCAT} \citep{Taylor}.}

%%%%%%%%%%%%%%%%%%%%%%%%%%
%      References
%%%%%%%%%%%%%%%%%%%%%%%%%%

\end{document}